\documentclass[10pt,twocolumn,twoside]{IEEEtran}

\newtheorem{theorem}{Theorem}
\newtheorem{lemma}{Lemma}
\newtheorem{definition}{Definition}
\newtheorem{proposition}{Proposition}
\newtheorem{corollary}{Corollary}

\usepackage{amsmath}
\usepackage{amssymb}
\usepackage{graphicx}

\ifCLASSOPTIONcompsoc
  \usepackage[nocompress]{cite}
\else
  \usepackage{cite}
\fi
\ifCLASSINFOpdf
\else
\fi
\begin{document}
%
\title{Adaptive Mitigation of Multi-Virus Propagation: A Passivity-Based Approach}
%
%
%
%

\author{Phillip Lee,~\IEEEmembership{Student Member,~IEEE,}
        Andrew Clark,~\IEEEmembership{Member,~IEEE,}
        Basel Alomair,~\IEEEmembership{Member,~IEEE,}
        Linda Bushnell,~\IEEEmembership{Senior Member,~IEEE,}
        and Radha Poovendran,~\IEEEmembership{Fellow,~IEEE}
\IEEEcompsocitemizethanks{\IEEEcompsocthanksitem P. Lee, L. Bushnell, and R. Poovendran are with the Network Security Lab, Department of Electrical Engineering, University of Washington, Seattle, WA 98195-2500. Email: \{leep3,lb2,rp3\}@uw.edu.
\IEEEcompsocthanksitem A. Clark is with the Department of Electrical and Computer Engineering, Worcester Polytechnic Institute, Worcester, MA 01609. Email: aclark@wpi.edu
\IEEEcompsocthanksitem B. Alomair is with the National Center for Cybersecurity Technology, King Abdulaziz City for Science and Technology, Riyadh, Saudi Arabia. Email:alomair@kacst.edu.sa}}
\IEEEtitleabstractindextext{%
\begin{abstract}
Malware propagation poses a growing threat to networked systems such as computer networks and cyber-physical systems. Current approaches to defending against malware propagation are based on patching or filtering susceptible nodes at a fixed rate. When the propagation dynamics are unknown or uncertain, however, the static rate that is chosen may be either insufficient to remove all viruses or too high, incurring additional performance cost. In this paper, we formulate adaptive strategies for mitigating multiple malware epidemics when the propagation rate is unknown, using patching and filtering-based defense mechanisms. In order to identify conditions for ensuring that all viruses are asymptotically removed, we show that the malware propagation, patching, and filtering processes can be modeled as coupled passive dynamical systems. We prove that the patching rate required to remove all viruses is bounded above by the passivity index of the coupled system, and formulate the problem of selecting the minimum-cost mitigation strategy. Our results are evaluated through a numerical study.
\end{abstract}}

\maketitle

\IEEEdisplaynontitleabstractindextext

\section{Introduction}\label{sec:introduction}
The growing reliance on computer networks for communication creates a corresponding increase in the threat of computer malware. Computer malware is an application that infects and installs itself on a host, and then uses the resources of that host to attempt to infect other devices.   Infected hosts often form large botnets that are controlled by one or more malicious adversaries and used to mount attacks including denial of service and spam campaigns~\cite{gu2007bothunter}.  Malware has been growing in sophistication, with new attack vectors targeting social networks~\cite{yan2011malware} and mobile devices~\cite{zhou2012dissecting}.

A variety of defense mechanisms have been developed for thwarting the spread of malware.  The standard approach is to periodically patch hosts against known malware, thus removing the infection and, depending on the type of malware, preventing reinfection in the future.  Proactive defenses include scanning network traffic with intrusion detection systems to identify malware signatures and quarantine infected hosts~\cite{zou2003worm}.  

While each defense mechanism mitigates the spread of malware, there is also an associated performance cost, including host downtime during patching, delays due to packet filtering, and allocation of system resources to decoy networks.  In order to determine appropriate parameters (e.g., patching rate) of a mitigation strategy that balance removal of malware with system performance, propagation models have been proposed that describe the rate of malware propagation, the impact of the attack, and the effectiveness of mitigation~\cite{zou2002code,bloem2009optimal}.  These models provide an analytical framework for designing a malware defense strategy.

Standard malware propagation models are based on epidemic dynamics such as Susceptible-Infected-Susceptible (SIS), which depend on the network topology, scanning rate of the malware, and probability that a scanned host becomes infected.  In general, however, propagation characteristics such as the scanning rate are unknown \emph{a priori}, leading to uncertainties in the design of mitigation parameters.  Such a mitigation strategy could incur unnecessarily large overhead or fail to control the spread of malware \cite{han2014data}.

 These uncertainties are especially pronounced when multiple malware strains propagate through a network simultaneously.  The interactions between different strains are complex and inherently unpredictable.  In the case of \emph{competing} malware, one malware strain may install anti-virus software in order to remove or block other malware from compromising the same host~\cite{bailey2005blaster}.  \emph{Co-existing} or \emph{colluding} malware, in contrast, may reside together on a single host, and the presence of one malware can facilitate other infections, e.g., by disabling firewalls and anti-virus.  At present, however, defense mechanisms that incorporate uncertainties in the propagation of a single malware, let alone multiple co-existing or competing malwares, are in the early stages.

In this paper, we develop a passivity-based approach to modeling and mitigating multiple malware propagations, using both static and adaptive defenses. 
By modeling the multi-virus propagation, patching, and filtering as \emph{passive dynamical systems}, we develop intuitive rules for updating the probability of packet inspection in order to guarantee removal of the viruses while minimizing performance overhead.    Our specific contributions are as follows:
 \begin{itemize}
\item We develop a passivity framework for modeling multi-virus propagation and mitigation under SIS malware propagation dynamics. We derive mean-field dynamical models of multi-virus propagation Markov process and  prove that the multi-virus propagation and mitigation can be viewed as coupled passive dynamical systems, and show that the required patching rate is characterized by the passivity index of the system. In the case when the propagation rates are known to the defender, we formulate the convex optimization problem of selecting the minimum-cost mitigation strategy to remove multiple viruses at a desired rate.


\item When the propagation rates are \emph{not known to the defender apriori}, we consider the class of adaptive patching and filtering based defenses. We propose two adaptive patching-based defenses. In the first defense, we derive an update rule that is guaranteed to ensure asymptotic removal of all viruses in this network. In the second defense, we derive a rule that can drive the probability of infection to be arbitrarily low in the single-virus case while minimizing the performance overhead of mitigation. 
    

 \item We analyze two performance characteristics of our patching and filtering strategies, namely the convergence rate of the network to the state where all viruses are removed, and the total cost of mitigation. We derive bounds on both characteristics as functions of the update parameters.

 \item We evaluate our approach via a numerical study. We numerically verify the accuracy of the mean-field approximation by comparing it to the underlying Markov stochastic model via Monte-Carlo method. In addition, we compare the convergence rates under coexisting and competing malware propagation and verify convergence of the adaptive patching and filtering dynamics to the desired steady-state.
 \end{itemize}

The paper is organized as follows.  Section \ref{sec:related} discusses the related work. Section \ref{sec:model} presents the adversary and defense models, and gives background on passivity.  Section \ref{sec:dynamics} introduces a dynamical model for the multi-virus propagation and mitigation, and presents our passivity-based approach to selecting a fixed patching rate when the propagation rate is known. Section \ref{sec:patching} describes and analyzes our proposed adaptive patching  strategies.  Section \ref{sec:filtering} discusses adaptive packet filtering strategies.  Section \ref{sec:simulation} presents simulation results.  Section \ref{sec:conclusion} concludes the paper. 
\section{Related Work}
\label{sec:related}
Malware propagation models have received significant research attention in recent years.  Standard approaches for modeling propagation of a single malware are based on ordinary differential equation models from epidemiology, such as the Kermack-McKendrick model~\cite{kephart1991directed}.  These models have been extensively analyzed theoretically and empirically, including applications to specific outbreaks such as the Code Red and Slammer worms~\cite{zou2002code}.   Eigenvalue bounds on the rate of malware propagation, as well as the threshold rate for patching infected nodes in order to eliminate viruses, were presented in \cite{wang2003epidemic}.  Multi-virus propagation has also received recent study~\cite{xu2012stochastic,watkins2015optimal}.   Propagation models have been developed to capture features of specific application domains, including mobile phones~\cite{zhou2012dissecting} and social networks~\cite{yan2011malware}.  Control-theoretic techniques for designing optimal malware propagation and attack strategies were presented in \cite{khouzani2012maximum}.

Dynamical models of virus propagation provide an analytical framework for designing mitigation strategies.  An optimal control approach to mitigating a single virus is given in \cite{bloem2009optimal}.    Geometric programming techniques for selecting the least-costly patching and vaccination rates were developed in \cite{preciado2013optimal}. Defenses against malware propagation in time-varying networks were considered in \cite{ogura2015stability}.  Recently, an optimization approach to defense against epidemics with uncertain propagation parameters was proposed~\cite{han2014data}.  Under this approach, fixed mitigation parameters were selected to ensure robustness to propagation parameters within an \emph{a priori} known range.  Our approach, on the other hand, adaptively increases the level of filtering in the network and makes no assumptions regarding the propagation parameters. An adaptive approach for virus mitigation under budget constraints was presented in \cite{drakopoulos2014efficient}.

The preliminary conference version of this work \cite{lee2015passivity} presented a passivity-based approach to modeling and mitigating multiple viruses using patching-based defenses with a fixed rate. The focus of \cite{lee2015passivity} was design of mitigation strategies to remove all viruses at a desired rate when the propagation rates are known to the defender.  It did not, however, consider the case when the propagation rates are unknown to the defender apriori. In addition, it only considered the impact of patching-based defense mechanism but not the filtering-based defense.
 
\section{Model and Preliminaries}
\label{sec:model}
This section presents the model and assumptions of the adversary and network defense.  We also give a brief background on passivity.

\subsection{Adversary Model}
\label{subsec:adversary}
We consider an undirected network with $N$ hosts. We say there exists an edge $(i,j)$ if hosts $i, j\in N$ can directly communicate with each other. The set of edges are denoted as $E$. A host $j$ is a neighboring host of ${i}$ if there exists an edge $(i,j)$ between ${i}$ and ${j}$. The set of neighboring hosts of host ${i}$ is denoted as $N_i$. Given a network topology, we define the adjacency matrix $A$ as $|N|\times |N|$ matrix with 0 on the diagonal entries and $A_{ij}=1$ if $(i,j)\in E$ and $A_{ij} = 0$ otherwise for the off-diagonal entries.

A set of malwares $V$ attempts to infect network hosts.  Once a host has been compromised by malware $v \in V$, that host will send malware traffic (e.g., embedded in email, social media, or other data flows) to non-infected neighboring hosts. 
We model the arrival process of malware traffic of virus $v$ as a Poisson process with rate $\mu^{v}$. In other words, the interarrival times of malware traffic are independent exponential random variables with mean $\frac{1}{\mu^{v}}$.
The receiving host becomes infected by each malware packet with  probability $p^{S,v}$, depending on the set of malwares $S$ currently infecting that host. This dependence is due to the fact that malwares may either install or disable anti-virus software onto a host, thus changing the difficulty of re-infection by a different malware.

A pair of malwares $v$ and $w$ can either be \emph{co-existing} or \emph{competing}.  If $v$ and $w$ are co-existing, then both can be present on the same host at any time.  If $v$ and $w$ are competing, then malware $v$ will attempt to remove malware $w$ if it is successfully installed on a host; hence, malwares $v$ and $w$ will never reside on the same host.  We let $C_{v}$ denote the set of malwares that compete with malware $v$.

\subsection{Network Defense Model}
\label{subsec:defense}
We consider two types of defense mechanisms, namely, \emph{patching} and \emph{packet filtering}. In the patching-based defense, each host is taken offline according to a random process and is inspected for any potential infection. When an infection is detected, the system administrator removes the infection and brings the host back online. The drawback of this defense mechanism is it could induce unnecessary cost of taking hosts offline since the patching process will continue even when all malwares are removed from the network since the inspection and cleaning process is independent from the state of the hosts. On the other hand, in the filtering-based mitigation, each packet that is sent from one host to the other is randomly forwarded according to an independent Bernoulli process to an \emph{intrusion detection system} (IDS), which inspects the packet for malware signatures.  If such signatures are detected, all malwares are removed from the host that sent the malware packet. Since a host is taken offline only when a packet that contains malware is detected, filtering-based mitigation avoids unnecessary cost. However, since the infected host will only send malware traffic to uninfected hosts, the filtering will not be able to detect any infection when all hosts are infected.

In this paper, we consider a susceptible-infected-susceptible (SIS) model \cite{watkins2015deterministic},where patched hosts can be reinfected at a later time.  In the patching defense, each host $i$ is taken offline according to a Poisson process with rate $\beta_{i}$, and patched against \emph{all} known malwares. That is, the times between two consecutive patching for host $i$ are modeled as independent exponential random variables with mean $\frac{1}{\beta_{i}}$.   In the packet filtering defense, we assume each packet that is sent from host $i$ to host $j$ is randomly forwarded with probability $q$ to an IDS. The parameters $\beta_{i}$ and $q$ vary over time, and are assumed to be set by a centralized entity, which is notified when a malware packet or infected host is detected.

\subsection{Background on Passivity}
\label{subsec:passivity}
This section gives background on passivity.  All definitions can be found in \cite{khalil2002nonlinear}.

\begin{definition}
\label{def:passivity} A dynamical system represented by the state model $\dot{x}(t) = f(x(t),u(t)), \: y(t) = h(x(t),u(t))$, where $f: R^{n} \times R^{p} \to R^{n}$ is locally Lipschitz, and $y: R^{n} \times R^{p} \to R^{p}$ is continuous, is called \emph{passive} if there exists a continuously differentiable positive semidefinite function $W$ (storage function) such that 
\begin{equation}
\dot{W}(t)\leq u^{T}y.
\end{equation}
If there exists a parameter $\rho$ such that 
\begin{equation}
\dot{W}(t)\leq \rho y^{T}y+ u^{T}y
\label{eq:ofp}
\end{equation}
for all $t$, then the system is called \emph{output feedback passive} (OFP). 
\end{definition}
 A subclass of output feedback passive systems is \emph{output strictly passive} systems. A dynamical system is output strictly passive if $\rho < 0$. The smallest $\rho$ that satisfies the condition (\ref{eq:ofp}) is defined as the output feedback passivity index of the system. If there exists a symmetric matrix $Q$ such that $\dot{W}(t) \leq y(t)^{T}Qy(t) + u(t)^{T}y(t)$ then the output feedback passivity index $\rho$ is upper bounded by $\rho \leq \mu_{1}(Q)$, where $\mu_{1}$ denotes the largest eigenvalue of $Q$.

  \begin{definition}
  \label{def:exp_stable}
  Given a dynamical system $\dot{\mathbf{x}}(t) = f(\mathbf{x}(t))$ with $\mathbf{x}(0) = \mathbf{x}_{0}$, where $f: D\to R^{n}$ for the domain $D \subset R^{n}$, is a locally Lipschitz function in $x$, an equilibrium point $\mathbf{x}^{*}$ is exponentially stable with rate of convergence $\alpha$ if there exist positive constants $c$ and $\alpha$ such that $$||\mathbf{x}(t)-\mathbf{x}^{*}|| \leq c \exp{(-\alpha t)}||\mathbf{x}_{0}||$$ for all initial states $\mathbf{x}_{0} \in D$.
  \end{definition}
  
   The following theorem gives a condition for exponential stability as well as bounds on the parameters $c$ and $\alpha$.
   \begin{theorem}
   \label{theorem:passivity_exp_stable}
   Let $\dot{\mathbf{x}}(t) = f(\mathbf{x}(t))$ be a dynamical system with equilibrium point $\mathbf{x}^{*}$.  Suppose that there exists a positive semidefinite function $W$ such that $W({\mathbf{x}^{*}}) = 0$ and positive constants $c_{1}$, $c_{2}$, $c_{3}$, and $p$ such that
   \begin{eqnarray*}
   c_{1}||\mathbf{x} - \mathbf{x}^{*}||^{p} \leq  &W(\mathbf{x})& \leq c_{2}||\mathbf{x} - \mathbf{x}^{*}||^{p}, \\
   \dot{W} &\leq& -c_{3}||\mathbf{x}-{\mathbf{x}^{*}}||^{p}.
   \end{eqnarray*}
   Then $\mathbf{x}^{*}$ is exponentially stable with $||\mathbf{x}(t) - \mathbf{x}^{*}||$ upper bounded by $$||\mathbf{x}(t) - \mathbf{x}^{*}|| \leq \left(\frac{c_{2}}{c_{1}}\right)^{1/p}\exp{\left(-\frac{c_{3}}{pc_{1}}t\right)}||\mathbf{x}_{0}||.$$
    \end{theorem} 
 The following two theorems provide sufficient conditions for asymptotic convergence. 
 \begin{theorem} \cite{khalil2002nonlinear} The negative feedback interconnection of two strictly passive systems $\dot{W}_{1}\leq u_{1}^{T}y_{1}$ and $\dot{W}_{2} \leq u_{2}^{T}y_{2}$ where $u_{2} = y_{1}$ and $u_{1} = -y_{2}$ is asymptotically stable.
 \label{thm:negative}
 \end{theorem}
 The following corollary follows directly from Definition \ref{def:passivity} and Theorem \ref{theorem:passivity_exp_stable}. 
 \begin{corollary} 
 \label{cor:exp}
 \cite{khalil2002nonlinear} Let $\dot{x}(t) = f(x(t), u(t))$, $y(t) = x(t)$ be an OFP system with equilibrium point $\mathbf{x}^{*} = 0$ admitting a quadratic storage function $W(\mathbf{x}) = \frac{1}{2} \mathbf{x}^{T}\mathbf{x}$. Let the OFP passivity index be $\rho$. Then $u = -(\rho+\epsilon)x$ will guarantee exponential stability with convergence rate $\epsilon$.   
 \end{corollary}

 \begin{theorem} 
 \label{thm:lasalle}\cite{khalil2002nonlinear} \emph{LaSalle's Invariance Principle}: Given a set $\Omega \subset D$ that is positively invariant with respect to dynamics $\dot{\mathbf{x}} = f(\mathbf{x})$, and $W: D \to R$ being a continuously differentiable function such that $\dot{W}(\mathbf{x}) < 0$ in $\Omega$, every solution starting in $\Omega$ will converge to the largest invariant set $M \subset \mathcal{I}$ where $\mathcal{I}$ is the set of points in $\Omega$ such that $\dot{W}(\mathbf{x}) = 0$.
 \end{theorem}

\section{Multi-Virus Propagation Dynamics}
\label{sec:dynamics}
In this section, a Markov model for malware propagation and mitigation is formulated.  A state-space dynamical model is derived using a mean-field approximation of the Markov propagation model.  We then prove that the propagation model is output feedback passive, as a first step towards a passivity-based approach to designing a mitigation strategy. We formulate the problem of selecting a static patching rate when the propagation parameters are known. 

\subsection{Markov Model and Mean-Field Approximation}
\label{subsec:Markov}
The time-varying components of the system model defined in Section \ref{sec:model} consist of the set of malwares infecting each host $i$ at time $t$, denoted $S_{i}(t) \subseteq V$, as well as the patching rate $\beta_{i}(t)$ of each host ${i}$ and the probability of packet filtering, denoted $q(t)$.  The quantities $\beta_{i}(t)$ and $q(t)$ vary over time due to the adaptive defense.  Taken together, $\mathcal{S}(t) = (S_{1}(t),\ldots,S_{n}(t), \beta_{1}(t),\ldots,\beta_{n}(t), q(t))$ comprises the state of the system.

Due to the Poisson assumption on the infection and patching rates, the state $\mathcal{S}(t)$ defines a continuous-time Markov chain with the following transition rates. For malware $v$, each infected host sends malware packets to each uninfected neighbor with rate $\mu^{v}$. When a host ${i}$ receives a packet infected with malware $v$ at time $t$, host ${i}$ becomes infected with malware $v$ and all competing viruses (i.e., $S_{i}(t) \cap C_{v}$) are removed with probability $p(S_{i}(t),v)$. Host $i$'s transition rate from being infected with a set of viruses $S_{i}(t)$ to being infected with $S_{i}(t) \setminus C_{v} \cup \{v\}$ due to a single neighbor infected with virus $v$ is denoted $\lambda^{S,v} \triangleq p(S,v)\mu^{v}$. Throughout this paper, we define $\lambda_{\max} = \max_{S,v} \lambda^{S,v}$ and $\lambda_{\min} = \min_{S,v} \lambda_{S,v}$.


Transitions due to the filtering process are described as follows.  For any malware $v$ with $v \in S_{i}(t) \setminus S_{j}(t)$ with $j \in N_{i}$, host ${i}$ sends malware packets to $j$ with rate $\mu^{v}$, which are inspected with probability $q(t)$.  If the malware packet is forwarded to IDS, then the host ${i}$ is taken offline and all malwares in $S_{i}(t)$ are removed, resulting in a transition from $S_{i}(t)$ to $\emptyset$ with rate $\overline{\lambda}^{v}(t) \triangleq q(t)\mu^{v}$. 
The last type of transition occurs due to the patching process.  This results in a transition from $S_{i}(t)$ to $\emptyset$ with rate $\beta_{i}(t)$.  

\begin{figure}[h]
\centering
\includegraphics[width=1.3in]{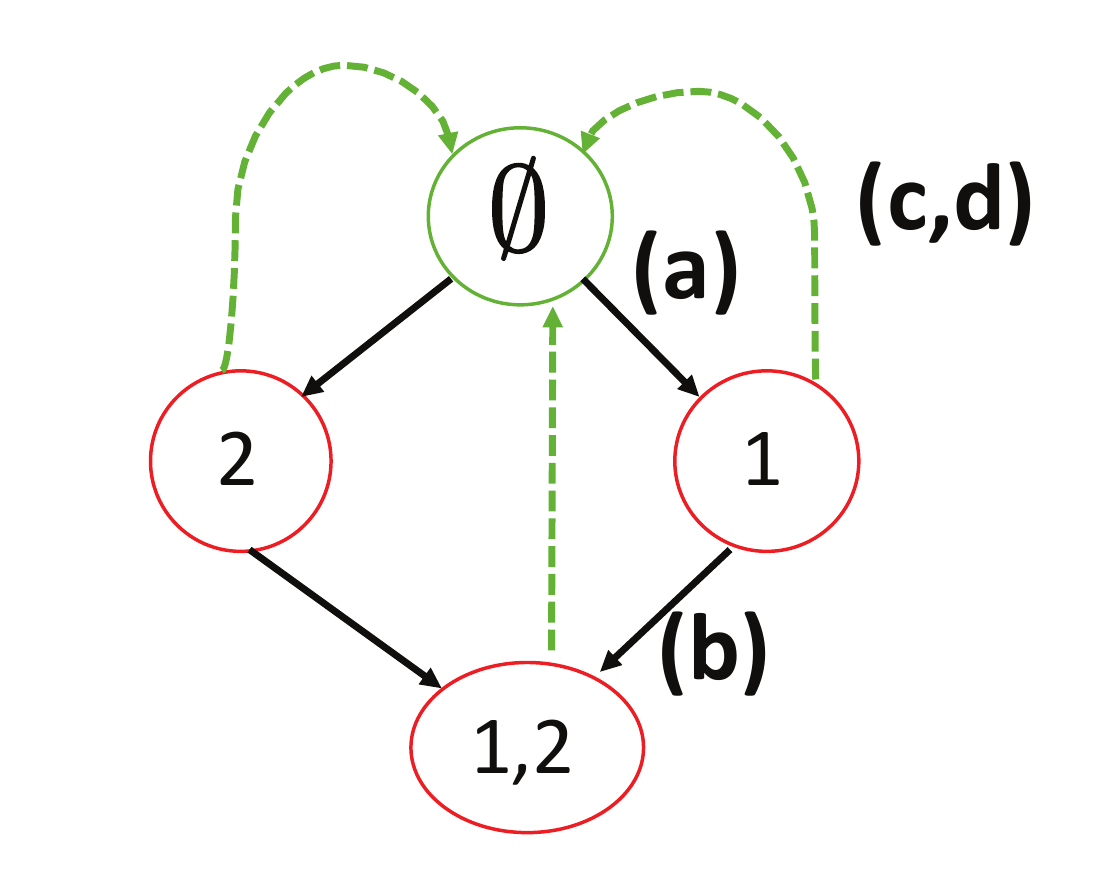}
\caption{Illustration of possible transitions with two malwares 1 and 2 that are coexisting. If $S = \{1\}$, then (a) is the transition into set $S$ by being infected with malware 1, (b) is the transition away from $S$ by being additionally infected with malware 2, and $(c), (d)$ is the transition away from set $S$ due to patching and filtering respectively.}
\label{fig:transition}
\end{figure}

The Markov model defined in this fashion has a number of states that is exponential in the number of hosts and malwares. Instead of dealing directly with all possible combinations of states $S_{i}(t)$, which is computationally infeasible for large networks, we consider the \emph{average} probability of infection for the tractability of analysis by applying mean-field approximation analogous to \cite{kephart1991directed,van2009virus,xu2012stochastic}.   The mean-field model is described by the states $\{x_{i}^{S}(t) : i \in N, S \subseteq V\}$, defined as the probability that host $i$ is infected with a set of viruses $S$ at time $t$. In describing the mean-field dynamics, we first observe that the set of subsets of $V$ that can transition to a set $S$ is given by $$\bigcup_{v \in S}{\{(S \setminus \{v\}) \cup R : R \subseteq C_{v}\}} \subset 2^{V}$$ where $2^{V}$ is the power set of the set $V$. Using the Kolmogorov forward equation \cite{ross2014introduction},  the net transitions into state $S_{i}$ are described by
\begin{IEEEeqnarray}{rCl}
\IEEEeqnarraymulticol{3}{l}{
\nonumber
\dot{x}_{i}^{S}(t)}\\
\label{eq:dynamics1}
&=& \sum_{v \in S}{\sum_{R \subseteq C_{v}}{\sum_{j \in N_{i}}{\sum_{T\subset V:T \ni v}{\left[\lambda^{(S \setminus \{v\} \cup R,v)}\right.}}}} \\
\nonumber
&& \times \left. Pr(S_{i}(t) = S \setminus \{v\} \cup R, S_{j}(t) = T)\right] \\
\label{eq:dynamics2}
&& - \sum_{v \notin S}{\sum_{j \in N_{i}}{\sum_{T \ni v}{\left[\lambda^{S,v}\right.}}} 
  \left. Pr(S_{i}(t) = S, S_{j}(t) = T)\right] \\
\label{eq:dynamics3}
&& - \sum_{v \in S}{\sum_{j \in N_{i}}{\sum_{T: v \notin T}{\overline{\lambda}^{v}(t)Pr(S_{i}(t) = S, S_{j}(t) = T)}}} \\
\label{eq:dynamics4}
&& - \beta_{i}(t)x_{i}^{S}(t).
\end{IEEEeqnarray}

In the above, Eq. (\ref{eq:dynamics1}) describes transitions to $S$ due to infection, while Eq. (\ref{eq:dynamics2}) describes transitions from $S$ due to infection with viruses not in $S$.  Eqs. (\ref{eq:dynamics3}) and (\ref{eq:dynamics4}) describe the impact of filtering and patching, respectively (Transitions (a), (b) and (c,d) respectively in Figure \ref{fig:transition}).

\subsubsection{Independence Approximation and its Implication} Throughout this paper, we make
an independence assumption that $Pr(S_{i}(t) = S, S_{j}(t) = T) = x_{i}^{S}x_{j}^{T}$ for all $i$, $j$, $S$, and $T$. 
With this assumption, the dynamics of $x_{i}^{S}(t)$ are rewritten as
\begin{IEEEeqnarray}{rCl}
\label{eq:indep_dynamics1}
&&\dot{x}_{i}^{S}(t) =\sum_{v \in S}{\sum_{R \subseteq C_{v}}{\sum_{j \in N_{i}}{\sum_{T \ni v}{\lambda^{(S \setminus \{v\} \cup R,v)}x_{i}^{S \setminus \{v\} \cup R}x_{j}^{T}}}}}\\
\label{eq:indep_dynamics2}
&& - \sum_{v \notin S}{\sum_{j \in N_{i}}{\sum_{T \ni v}{\lambda^{S,v}x_{i}^{S}x_{j}^{T}}}} \\
\label{eq:indep_dynamics3}
&& - \sum_{v \in S}{\sum_{j \in N_{i}}{\sum_{T: v \notin T}{\overline{\lambda}^{v}(t)x_{i}^{S}x_{j}^{T}}}} 
\label{eq:indep_dynamics4}
 - \beta_{i}(t)x_{i}^{S}(t).
\end{IEEEeqnarray}
This independence assumption is common in models of malware propagation \cite{van2009virus,watkins2015deterministic}. In the case of single virus propagation, this assumption is known to overestimate the mean-field propagation dynamics \cite{van2009virus} under the assumption    
\begin{equation}
Pr(S_{i} = \emptyset |S_{j} = \{v\})\leq Pr(S_{i} = \emptyset).
\label{eq:ind}
\end{equation}
In other words, conditioned on the event that a neighboring host $j$ of host ${i}$ is infected, it cannot increase the probability that ${i}$ is clean. Since the independence assumption overestimates the propagation dynamics, it implies that any mitigation strategy that is sufficient to remove all malwares with the independence assumption is also sufficient to remove all malwares for the underlying mean-field dynamics. Recently, in the case of competing multi-virus propagation, it was shown \cite{watkins2015deterministic} that the independence assumption does not systematically over or under estimate the propagation dynamics of \emph{individual} malware propagation (equations (\ref{eq:indep_dynamics1}) and (\ref{eq:indep_dynamics2})). On the other hand, if the goal of the defender is to remove \emph{all} malwares, not the individual malware, then it suffices to consider the dynamics of $\bar{x}_{i}(t) = \sum_{S\subset V:S\neq \emptyset} x_{i}^{S}$, the probability that host $i$ is infected with at least one malware at time $t$. The following theorem shows that the conditional probability assumption (\ref{eq:ind}) results in over-estimation of mean field dynamics of $\bar{x}_{i}(t)$. 
\begin{figure}
\centering
\includegraphics[width=2.7in]{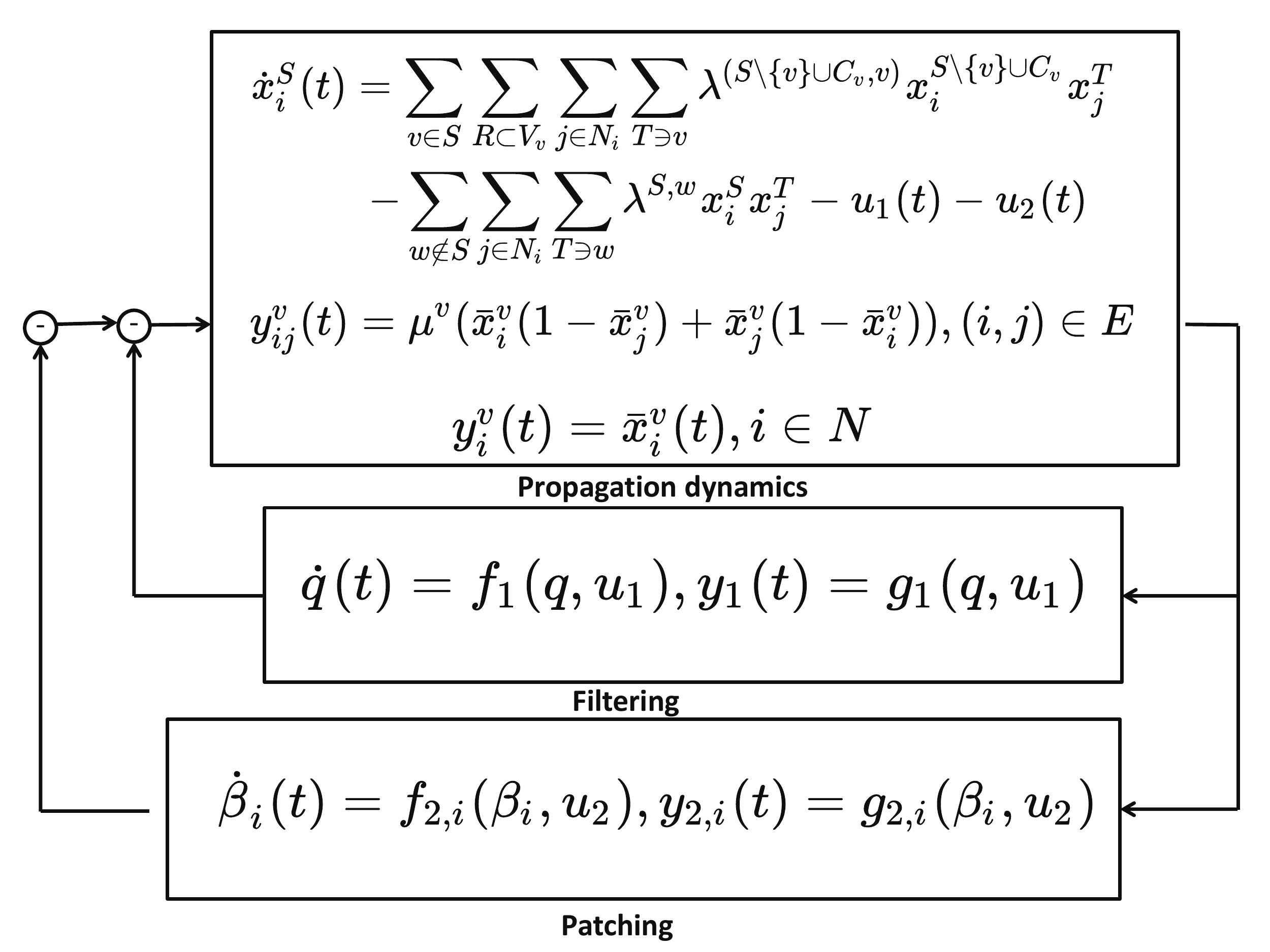}
\caption{Representation of our passivity-based approach, consisting of coupled dynamical systems representing propagation, filtering, and patching.}
\label{fig:representation}
\end{figure}
\begin{theorem} 
\label{thm:xbar}
Consider the propagation dynamics of $\bar{x}_{i}(t)$ in the absence of mitigation strategy given as 
\begin{equation}
\dot{\overline{x}}_{i} = (1-\overline{x}_{i})\sum_{j \in N_{i}}{\sum_{v \in V}{\lambda^{\emptyset,v}\:\overline{x}_{j}^{v}}}.
\label{eq:xbar}
\end{equation}
The dynamics (\ref{eq:xbar}) provides an upperbound on the mean-field dynamics of $\bar{x}_{i}$.
\end{theorem}
\begin{IEEEproof} Define $\bar{x}_{i}^{v} = \sum_{S \ni v} x_{i}^{S}$, the probability that host $i$ is infected with malware $v$ at time $t$, and  $\gamma_{i}^{S\to S^{'}}(t)$ as the transition rate of being infected with set of viruses $S^{'}$ from being infected with $S$. $\mathbf{1}_{S_{i}(t) = S}$ is an indicator which equals to 1 if node ${i}$ is infected with set of viruses $S$ and 0 otherwise. Then from equations (\ref{eq:indep_dynamics1}), (\ref{eq:indep_dynamics2}), we have
\begin{eqnarray*}
\dot{\bar{x}}_{i}(t)
&=&  \sum_{S\subset V: S\neq \emptyset} \mathbb{E}[\sum_{S^{'} \neq S} \mathbf{1}_{S_{i}(t) = S^{'}} \gamma_{i}^{S^{'}\to S} (t)]\\
&-& \mathbb{E}[\sum_{S^{'} \neq S: S^{'}\neq \emptyset} \mathbf{1}_{S_{i}(t) = S} \gamma_{i}^{S\to S^{'}} (t)]\\
&=&  \sum_{v \in V}  \mathbb{E}[\mathbf{1}_{S_{i}(t) = \emptyset} \sum_{j\in N_{i}} \sum_{S\ni v} \lambda^{\emptyset, v} \mathbf{1}_{S_{j}(t) = S}] \\
&=& \mathbb{E}[\mathbf{1}_{S_{i}(t) = \emptyset} \sum_{j \in N_{i}} \sum_{v\in V} \sum_{S\ni v} \lambda^{\emptyset,v} \mathbf{1}_{S_{j}(t) = S}]  \\
& \leq & (1-\bar{x}_{i}(t)) \sum_{j\in N_{i}} \sum_{v\in V} \lambda^{\emptyset,v}\bar{x}_{j}^{v}. 
\end{eqnarray*} 
where the last inequality is from the assumption (\ref{eq:ind}).
\end{IEEEproof}

 In Section \ref{sec:simulation}, we empirically analyze the accuracy of our approximate dynamical model showing that the mean-field dynamics closely follow the underlying Markov process.

\subsection{Passivity Analysis of Malware Propagation}
\label{subsec:passivity_SIS}
Our passivity-based analysis of malware propagation and mitigation decomposes the propagation model into three coupled dynamical systems, namely, multi-virus propagation, filtering-based mitigation, and patching based mitigation (Figure \ref{fig:representation}). The first step in developing our approach is to prove that the propagation dynamics (top block) are output feedback passive.

As a preliminary, we have the following result that provides a storage function for the set of dynamical systems from the mean-field dynamics.

\begin{lemma}
\label{lemma:passivity_SIS}
Consider a finite set $V$, and the set of state dynamics given as 
\begin{equation*}
\dot{x}^{S} = -\sum_{T \neq S} \gamma_{S \to T}(t) x^{S}(t) +  \sum_{T \neq S} \gamma_{T \to S}(t) x^{T}(t)
\end{equation*}
for $S, T\subset V$.
Given a quadratic function $W = \frac{1}{2} \sum_{ S\subset V} (x^{S})^{2}$, we have 
\begin{equation*}
\dot{W} = \sum_{T\neq S} \sum_{S \subset V} \gamma_{S \to T}(t) \left(-(x^{S})^{2}(t) + x^{S}(t) x^{T}(t)\right).
\end{equation*}
\end{lemma}
\begin{IEEEproof}

\begin{eqnarray*}
\dot{W} &=& \sum_{S} x^{S} \dot{x}_{S}\\
&=& \sum_{S } \left(-\sum_{T \neq S}\gamma_{S\to T}(t) (x^{S})^{2}  + \sum_{T\neq S}\gamma_{T \to S}(t) x^{S} x^{T}\right)\\
&=& \sum_{T\neq S} \sum_{S \subset V } \gamma_{S \to T}(t) \left(-(x^{S})^{2}(t) + x^{S}(t) x^{T}(t)\right)
\end{eqnarray*}

\end{IEEEproof}

In what follows, we analyze the storage function $$W_{i}(\mathbf{x}) = \frac{1}{2}\sum_{S \neq \emptyset}{(x_{i}^{S})^{2}}$$ using the results of Lemma \ref{lemma:passivity_SIS}. 
\begin{lemma}
\label{lemma:passivity_SIS_2} Define $\mathbf{u}_{i} = -\beta_{i} \mathbf{x}_{i}$, and $\mathbf{x}_{i}$ as a column vector of length $2^{|V|}-1$ where entries enumerate $\{x_{i}^{S}\}$ for all possible subset $S \subset V \setminus \emptyset$. The time derivative of $W_{i}(\mathbf{x})$ is given by
\begin{displaymath}
\dot{W}_{i} \leq \mathbf{x}_{i}^{T}Q_{i}\mathbf{x}_{i} + \sum_{j \in N_{i}}{\mathbf{x}_{j}^{T}Q\mathbf{x}_{j}} + \mathbf{u}_{i}^{T}\mathbf{x}_{i},
\end{displaymath}
where $Q_{i}$ is a diagonal matrix with diagonal entry corresponding to host $i$'s state being $S$. The diagonal entries of $Q_{i}$ are written as $$Q_{i}(S,S) = \frac{|N_i|}{6}\sum_{v \in S}{\sum_{R \subseteq C_{v}}{2^{|V \setminus C_{v}|-1}\lambda^{S \setminus \{v\} \cup R, v}}}$$ and $Q = H\Lambda H^{T}$. Here $H$ is a $2^{|V|} \times |V|$ 0-1 matrix where each entry $H_{S,v}$ corresponds to set $S$ (row) and malware $v$ (column), which equals to 1   
if $v \in S$ and 0 otherwise, and $\Lambda$ is a $|V| \times |V|$ diagonal matrix with $$\Lambda_{vv} = \frac{1}{12}\sum_{S: v \notin S}{\lambda^{S,v}}.$$    
\end{lemma}

\begin{IEEEproof}
Let $\mathcal{R}$ be the set of realizable sets where for $S\in \mathcal{R}$, if $v\in S$ then for any $u\in C_{v}$, $u\not \in S$.
By Lemma \ref{lemma:passivity_SIS}, $\dot{W}_{i}(\mathbf{x})$ is equal to 
\begin{IEEEeqnarray*}{rCl}
\dot{W}_{i} &=& \sum_{S \in \mathcal{R}}{\sum_{v \in S}{\sum_{R \subseteq C_{v}}{\left[\gamma_{S \setminus \{v\} \cup R \rightarrow S}(t)(-(x_{i}^{S \setminus \{v\} \cup R})^{2} \right.}}} \\
&& \left. + x_{i}^{S \setminus \{v\} \cup R}x_{i}^{S})\right] \\
&\leq& \frac{1}{4}\sum_{S \in \mathcal{R}}{\sum_{v \in S}{\sum_{R \subseteq C_{v}}{\left[\gamma_{S \setminus \{v\} \cup R \rightarrow S}(t)(x_{i}^{S})^{2}\right]}}},
\end{IEEEeqnarray*}
where the inequality follows from the identity $(2x_{i}^{S \setminus \{v\} \cup R} - x_{i}^{S})^{2} \geq 0$.
Since $\gamma_{S \setminus \{v\} \cup R \rightarrow S}(t) = \sum_{n_j \in N_{i}}{\lambda^{S \setminus \{v\} \cup R, v}\overline{x}_{j}^{v}},$ we have 
\begin{IEEEeqnarray*}{rCl}
\dot{W}_{i} &\leq& \frac{1}{4}\sum_{S \in \mathcal{R}}{\sum_{v \in S}{\sum_{R \subseteq C_{v}}{\sum_{n_j \in N_{i}}{\lambda^{S \setminus \{v\} \cup R,v} \:\overline{x}_{j}^{v}(x_{i}^{S})^{2}}}}} \\
&\leq& \frac{1}{12}\sum_{S \in \mathcal{R}}{\sum_{v \in S}{\sum_{R \subseteq C_{v}}{\sum_{j \in N_{i}}{\lambda^{S \setminus \{v\}\cup R, v}\:(\overline{x}_{j}^{v})^{2}}}}} \\
&& + \frac{1}{6}\sum_{S \in \mathcal{R}}{\sum_{v \in S}{\sum_{R \subseteq C_{v}}{\sum_{j \in N_{i}}{\lambda^{S \setminus \{v\} \cup R,v}\:(x_{i}^{S})^{2}}}}} 
\end{IEEEeqnarray*}
by the inequality $abc \leq \frac{1}{3}(a^{2}+b^{2}+c^{2})$. We can simplify the the first term as follows:
\begin{eqnarray*}
&&\sum_{S \in \mathcal{R}}{\sum_{v \in S}{\sum_{R \subseteq C_{v}}{\sum_{j \in N_{i}}{\lambda^{S \setminus \{v\} \cup R,v}\:(\overline{x}_{j}^{v})^{2}}}}}\\
&=& \sum_{v \in S}{\sum_{\stackrel{S \in \mathcal{R}:}{v \in S}}{\sum_{R \subseteq C_{v}}{\sum_{j \in N_{i}}{\lambda^{S \setminus \{v\} \cup R,v}\:(\overline{x}_{j}^{v})^{2}}}}} \\
&=& \sum_{v \in S}{\sum_{j \in N_{i}}{\left[(\overline{x}_{j}^{v})^{2}\left(\sum_{{S \in \mathcal{R}:}{v \notin S}}{\lambda^{S \setminus \{v\},S}}\right)\right]}}.
\end{eqnarray*}
The last equivalence relationship follows from the observation that each set $T \in \mathcal{R}$ with $v \notin T$ appears exactly once in the collection $\mathcal{C} = \{(S \setminus \{v\} \cup R: S \in \mathcal{R}, R \subseteq C_{v}\}$. To see this, let $T \in \mathcal{R}$ be a set with $v \notin T$. Let $S = (T \setminus C_{v}) \cup \{v\}$ and $R = T \cap C_{v}$. We have $T = S \cup R \setminus \{v\}$, which appears in the collection $\mathcal{C}$. 

Now, suppose that there exist $S^{\prime}$ and $R^{\prime}$ with $v \in S^{\prime}$, $S^{\prime} \in \mathcal{R}$, $R^{\prime} \subseteq C_{v}$, $T = S^{\prime} \cup R^{\prime} \setminus \{v\}$, and $S^{\prime} \neq S$ or $R^{\prime} \neq R$. We have four cases. First, if $S^{\prime} \neq S$ and there exists $u \in S^{\prime} \setminus S$, then we must have $u \in R$. However, $u \in R \subseteq C_{v}$, and hence $u$ and $v$ are both in $S^{\prime}$, contradicting the assumption that $S^{\prime} \in \mathcal{R}$.

Second, suppose that $u \in S \setminus S^{\prime}$. By a similar argument, we must have $u^{\prime} \in R^{\prime} \subseteq C_{v}$, creating a contradiction by the same argument.

Third, suppose that there exists $u \in R^{\prime} \setminus R$. We must have $u \in S$, however, $u \in C_{v}$, creating a contradiction since $u,v \in S$ and $S \in \mathcal{R}$. Finally, the case where there exists $u \in R \setminus R^{\prime}$ is similar. 

This yields
\begin{multline*}
\dot{W}_{i}
\leq \frac{1}{12}\sum_{j \in N_{i}}{\sum_{v \in V}{\left[\left(\sum_{S: v \notin S}{\lambda^{S,v}}\right)\left(\sum_{T: v \in T}{x_{j}^{T}}\right)^{2}\right]}} \\
 + \frac{|N_{i}|}{6}\sum_{S \in \mathcal{R}}{\left(\sum_{v \in S}{\sum_{R \subseteq C_{v}}{\lambda^{S \setminus \{v\} \cup R,v}}}\right)(x_{i}^{S})^{2}}.
\end{multline*}
\end{IEEEproof}

Two special cases are a set of \emph{competing viruses}, in which $C_{v} = V \setminus \{v\}$ for all $v \in V$, and \emph{coexisting viruses}, in which $C_{v} = \emptyset$ for all $v \in V$. In the coexisting virus case,
$$Q_{i}(S,S) = \frac{|N_{i}|}{6}\sum_{v \in S}{2^{|V|-1}\lambda^{S \setminus \{v\},v}},$$ while in the competing case $$Q_{i}(S,S) = \frac{|N_{i}|}{6}\left[\sum_{u \neq v}{\lambda^{u,v}} + \lambda^{\emptyset,v}\right].$$  In general, the passivity index in the competing case will be less than the passivity index in the coexisting case since the exponential term $2^{|V|-1}$ will increase exponentially as the number of malwares increase. 

The following theorem implies that the multi-virus propagation is output-feedback passive, and hence that passivity-based techniques can be developed to design a mitigation strategy from Corollary \ref{cor:exp}.
\begin{theorem}
\label{theorem:SIS_passive_2}
The mean-field approximation (\ref{eq:indep_dynamics1})--(\ref{eq:indep_dynamics4}) of the multi-virus propagation dynamics without filtering ($\bar{\lambda}^{v} = 0$) is output feedback passive from input $(\mathbf{u}_{i}=-\beta_{i}\mathbf{x}_{i} : i \in N)$ to output $(\mathbf{x}_{i} : i \in N)$, with passivity index $\rho$ bounded by 
$$\rho \leq \max_{i}\left\{{\mu_{1}(Q_{i} + |N_{i}|Q)}\right\},$$ where $\mu_{1}(\cdot)$ denotes the largest eigenvalue of a matrix. 
\end{theorem}

\begin{IEEEproof}
Select the storage function $W(\mathbf{x}) = \sum_{i \in N}{W_{i}(\mathbf{x})}$. By Lemma \ref{lemma:passivity_SIS_2}, 
\begin{IEEEeqnarray*}{rCl}
\dot{W}(\mathbf{x}) &=& \sum_{i \in N}{\dot{W}_{i}} \leq \sum_{i\in N}{\mathbf{x}_{i}^{T}Q_{i}\mathbf{x}_{i}} + \sum_{i}{\sum_{j\in N_{i}}{\mathbf{x}_{j}^{T}Q\mathbf{x}_{j}}} + \sum_{i\in N}\mathbf{u}_{i}^{T}\mathbf{x}_{i} \\
&=& \sum_{i\in N}{\mathbf{x}_{i}^{T}Q_{i}\mathbf{x}_{i}} + \sum_{i}{|N_{i}|\mathbf{x}_{i}^{T}Q\mathbf{x}_{i}} + \sum_{i\in N}\mathbf{u}_{i}^{T}\mathbf{x}_{i} \\
&=& \sum_{i\in N}{\mathbf{x}_{i}^{T}(Q_{i}+|N_{i}|Q)\mathbf{x}_{i}} +\sum_{i\in N}\mathbf{u}_{i}^{T}\mathbf{x}_{i},
\end{IEEEeqnarray*}
implying that the system is OFP with passivity index $\max_{i}{\{\mu_{1}(Q_{i} + |N_{i}|Q)\}}$. 
\end{IEEEproof}
This completes the first step of proving passivity of the propagation dynamics in order to design the patching strategy to remove all malwares at a desired rate.

\subsection{Design of Static Patching Strategies}
\label{subsec:static_mitigation}
If the compromise rates $\lambda^{S,v}$ are known for all $S$ and $v$, then the results of Lemma \ref{lemma:passivity_SIS_2} and Theorem \ref{theorem:SIS_passive_2} can be used to select the patching rates $\{\beta_{i} : i \in N\}$ while minimizing a desired cost function. The following proposition provides a sufficient condition for removal of all viruses at a desired rate $\epsilon$.

\begin{proposition}
\label{prop:static_mitigation}
Let $B_{i} = \beta_{i}I_{(2^{|V|}-1) \times (2^{|V|}-1)}$, where $I$ denotes the identity matrix, and let $B$ be a block diagonal matrix with the $B_{i}$'s as diagonal entries. Define $\overline{Q}$ by $$\overline{Q} = A \otimes Q + \left(
\begin{array}{ccc}
Q_{1} & \cdots & 0 \\
\vdots & \ddots & \vdots \\
0 & \cdots & Q_{n}
\end{array}
\right).$$
where $A$ is the adjacency matrix of the network, and $\otimes$ is Kronecker product. 
If $B-\overline{Q} \geq \epsilon I$, where ``$\geq$'' denotes inequality in the semidefinite cone, then all viruses will be removed in steady-state and $||\mathbf{x}(t)||_{2} \leq \sqrt{|N|} e^{-\epsilon t}$ for all $t \geq 0$.
\end{proposition}

\begin{IEEEproof}
Using the storage function $W(\mathbf{x}) = \frac{1}{2}\mathbf{x}^{T}\mathbf{x}$, Theorem \ref{theorem:SIS_passive_2} implies that 
\begin{displaymath}
\dot{W}(\mathbf{x}) \leq \mathbf{x}^{T}\overline{Q}\mathbf{x}^{T} - \mathbf{x}^{T}B\mathbf{x}^{T} \leq -\epsilon \mathbf{x}^{T}\mathbf{x}.
\end{displaymath}
Therefore, from Theorem \ref{theorem:passivity_exp_stable}, we have $$||\mathbf{x}(t)||_{2} \leq e^{-\epsilon t}||\mathbf{x}(0)||_{2} \leq \sqrt{|N|}e^{-\epsilon t},$$ since $\sum_{S}{x_{i}^{S}(0)} \leq 1$. 
\end{IEEEproof}

Proposition \ref{prop:static_mitigation} implies that an optimal patching strategy can be selected using semidefinite programming, with the problem formulation
\begin{equation}
\label{eq:static_mitigation_opt}
\begin{array}{cc}
\mbox{minimize} & \sum_{i \in N}{c_{i}(\beta_{i})} \\
\mbox{s.t.} & B \geq \overline{Q}+\epsilon I \\
& \beta_{i} \geq 0 \ \forall i,
\end{array}
\end{equation}
where the cost function of patching for host $i$, $c_{i}$ is an increasing, convex function in $\beta_{i}$. 

When the infection parameters $\lambda^{S,v}$ are known, the optimization problem (\ref{eq:static_mitigation_opt}) can be used to select an efficient mitigation strategy. In general, however, these parameters will be unknown. One approach to incorporating unknown infection rates is through robust variations on (\ref{eq:static_mitigation_opt}), which would select the minimum-cost mitigation strategy over a set of possible mitigation strategies. Alternatively, an adaptive approach can be designed that dynamically adjusts the patching rate based on previously observed infections. Developing such an approach is the focus of the next section.

\section{Patching-Based Adaptive Mitigation}
\label{sec:patching}
This section presents two adaptive strategies for tuning the patching rate based on previous detections of infected hosts. The convergence of the patching rate and infection probability are analyzed for both rules.


\subsection{Adaptive Patching Strategy}
\label{subsec:adaptive_patching}
As in the previous section, we take a passivity-based approach to designing the patching strategy; the approach, however, is based on an equivalent representation of the malware propagation dynamics with different input and output. We define the probability that host $i$ is infected with at least one malware at time $t$ as $\bar{x}_{i}(t)$ and the probability that host $i$ is infected with virus $v$ as $\bar{x}_{i}^{v} = \sum_{S \ni v} x_{i}^{S}$. We use the state dynamics of $\overline{x}_{i} = \sum_{S}{x_{i}^{S}}$ under the independence assumption derived in Theorem \ref{thm:xbar} as $$\dot{\overline{x}}_{i} = (1-\overline{x}_{i})\sum_{j \in N_{i}}{\sum_{v \in V}{\lambda^{\emptyset,v}\:\overline{x}_{j}^{v}}} - \beta_{i}\overline{x}_{i}.$$ 

\begin{proposition}
 \label{prop:H1_passive_patching}
The dynamics of $\overline{x}_{i}$  are passive from input $\mathbf{u}$ where $u_{i}= (|N_{i}|\hat{\lambda} - \beta_{i})$ to output $\mathbf{y}$ where $y_{i}=(\overline{x}_{i})^{2}$.
\end{proposition}
\begin{IEEEproof}
Define the storage function $W(\mathbf{x}) = \frac{1}{2}\sum_{i \in N}{(\overline{x}_{i})^{2}}$ and $\hat{\lambda} = \sum_{ v \in V}{\lambda^{\emptyset,v}}$.  Differentiating with respect to time gives
\begin{IEEEeqnarray*}{rCl}
\dot{W}(\mathbf{x}) &=& \sum_{i \in N}{\overline{x}_{i}(1-\overline{x}_{i})\sum_{v \in V}{\sum_{j \in N_{i}}{\lambda^{\emptyset,v}\:\overline{x}_{j}^{v}}}} - \sum_{i \in N}{\beta_{i}(t)(\overline{x}_{i})^{2}} \\
&\leq& \sum_{i \in N}{\overline{x}_{i}(1-\overline{x}_{i})\sum_{j \in N_{i}}{\hat{\lambda}\overline{x}_{j}}} - \sum_{i \in N}{\beta_{i}(\overline{x}_{i})^{2}} \\
&\leq& \sum_{i \in N}{\sum_{j \in N_{i}}{\hat{\lambda}\overline{x}_{i}\overline{x}_{j}}} - \sum_{i \in N}{\beta_{i}(\overline{x}_{i})^{2}} \\
&\leq& \sum_{(i,j) \in E}{\frac{\hat{\lambda}}{2}((\overline{x}_{i})^{2} + (\overline{x}_{j})^{2}} - \sum_{i \in N}{\beta_{i}(\overline{x}_{i})^{2}} \\
&=& \sum_{i \in N}{(|N_{i}|\hat{\lambda}-\beta_{i})(\overline{x}_{i})^{2}},
\end{IEEEeqnarray*}
thus proving passivity.
\end{IEEEproof}

The passivity of the propagation dynamics implies that, in order to ensure convergence to the state where all viruses are removed, it suffices to select an update rule $\dot{\beta}_{i}(t)$ that is passive from input $(\overline{x}_{i})^{2}$ to output $(|N_{i}|\hat{\lambda} - \beta_{i})$ by Theorem \ref{thm:negative}.  One such adaptive rule is given by 
\begin{equation}
\label{eq:adaptive_patching}
\dot{\beta}_{i}(t) = \alpha \overline{x}_{i}
\end{equation}
 for some $\alpha > 0$. This patching strategy can be implemented by incrementing the patching rate by $\frac{\alpha}{\beta_{i}(t)}$ when an infection is detected. This is because the rate of this patching update process is $\beta_{i}(t) \bar{x}_{i}(t)$ at time $t$, which leads to the rate of change in the patching rate being equal to $\frac{\alpha}{\beta_{i}(t)}\beta_{i}(t) \bar{x}_{i}(t) = \alpha x_{i}(t)$. The adaptive patching does not require the knowledge of the propagation rate $\lambda^{v}$, but requires that $\beta_{i}(0) > 0$.
 

\begin{theorem}
\label{theorem:patching_stable}
Under the patching update rule $\dot{\beta}_{i}(t) = \alpha \overline{x}_{i}(t)$, $\lim_{t \rightarrow \infty}{\overline{x}_{i}(t)} = 0$ for all $i \in N$, implying that all malwares are removed asymptotically from the network.
\end{theorem}

\begin{IEEEproof}
Let $\mathbf{x}$ be the vector enumerating $\bar{x}_{i}^{v}$ for all $i\in N$ and $v\in V$. The proof is via the LaSalle Invariance Principle (Theorem \ref{thm:lasalle}). Define the storage function $W(\mathbf{x},\mathbf{\beta})$ by $$W(\mathbf{x},\boldsymbol{\beta}) = \frac{1}{2}{\sum_{i \in N}{(\overline{x}_{i})^{2}}} + {\sum_{i \in N}{\Gamma_{i}(\beta_{i})}},$$ where 
\begin{displaymath}
\Gamma_{i}(\beta_{i}) = \left\{
\begin{array}{ll}
\frac{1}{2\alpha}(|N_{i}|\hat{\lambda}-\beta_{i})^{2}, & \beta_{i} \leq |N_{i}|\hat{\lambda} \\
0, & \mbox{else}.
\end{array}
\right.
\end{displaymath}
By inspection, $W$ is positive semidefinite, and continuously differentiable, and therefore is a valid storage function. We now show that $\dot{W}(\mathbf{x},\boldsymbol{\beta}) \leq 0$. By Proposition \ref{prop:H1_passive_patching}, 
\begin{IEEEeqnarray*}{rCl}
\dot{W}(\mathbf{x}, \boldsymbol{\beta}) &\leq& \sum_{i \in N}{(|N_{i}|\hat{\lambda}-\beta_{i})(\overline{x}_{i})^{2}} + \sum_{i \in N}{\dot{\Gamma}_{i}(\beta_{i})} \\
&=& \sum_{i \in N}\left[(|N_{i}|\hat{\lambda}-\beta_{i})(\overline{x}_{i}^{2} - \bar{x}_{i})\right].
\end{IEEEeqnarray*}
We show that each term of the inner summation is bounded above by zero. If $\beta_{i} \leq |N_{i}|\hat{\lambda}$, then, the corresponding term is given by  $$(|N_{i}|\hat{\lambda}-\beta_{i})((\overline{x}_{i})^{2} - \overline{x}_{i}) \leq 0,$$ since $(\overline{x}_{i})^{2} \leq \overline{x}_{i}$ for $\overline{x}_{i}^{v} \in [0,1]$. On the other hand, if $\beta_{i} > |N_{i}|\hat{\lambda}$, then the corresponding term is simply $(|N_{i}|\hat{\lambda}-\beta_{i})(\overline{x}_{i})^{2} \leq 0$. 

By the LaSalle's Invariance Principle, all trajectories of $(\mathbf{x},\boldsymbol{\beta})$ converge to $\{(\mathbf{x},\boldsymbol{\beta}) : \dot{W}(\mathbf{x},\boldsymbol{\beta}) = 0\}$. We show that this set is equal to $\{(\mathbf{x}, \boldsymbol{\beta}) : \mathbf{x} = 0\}$. Since
\begin{IEEEeqnarray*}{rCl}
\dot{W}(\mathbf{x}, \boldsymbol{\beta}) &=& \sum_{i \in N}{\overline{x}_{i}(1-\overline{x}_{i})\sum_{v \in V}{\sum_{j \in N_{i}}{\lambda^{\emptyset,v}\overline{x}_{j}^{v}}}} - \sum_{i \in N}{\beta_{i}(t)(\overline{x}_{i})^{2}}\\
 &+& \sum_{i\in N} \dot{\Gamma}_{i}(\beta_{i}),
\end{IEEEeqnarray*}
we have $\dot{W}(\mathbf{0}, \boldsymbol{\beta}) = \sum_{i\in N} \dot{\Gamma}_{i}(\beta_{i})$. However $\dot{\Gamma}_{i}(\beta_{i}) = -(|N_{i}|\hat{\lambda} - \beta_{i})\bar{x}_{i}$ for $\beta_{i} < |N_{i}|\hat{\lambda}$ and 0 for $\beta_{i} \geq |N_{i}|\hat{\lambda}$, resulting in $\dot{\Gamma}_{i}(\beta_{i})=0$ if $x_{i} = 0$. Moreover, suppose there exists $(\mathbf{x},\boldsymbol{\beta})$ such that $\dot{W}(\mathbf{x}, \boldsymbol(\beta)) = 0$ and $\bar{x}_{i}^{v} > 0$ for some $i$ and $v$. Since $\bar{x}_{i} \geq x_{i}^{v} > 0$, we have $\dot{\beta}_{i} = \alpha \bar{x}_{i} > 0$. Thus $\beta_{i}$ will increase at $(\mathbf{x}, \boldsymbol{\beta})$, and hence such $(\mathbf{x}, \boldsymbol{\beta})$ cannot stay in the set where $\dot{W} = 0$. Therefore, $\dot{W}(\mathbf{x}, \boldsymbol{\beta})$ is identically 0 if and only if $\mathbf{x} = 0$.

\end{IEEEproof}

\subsection{Adaptive Patching Rate Analysis}
\label{subsec:adaptive_patching}
We now analyze the time required for the adaptive patching rate to converge to $\beta_{i}(t) = |N_{i}|\hat{\lambda}$.  As an approximation, we assume that the malware propagation $\overline{x}_{i}^{v}(t)$ instantaneously converges to a fixed point, denoted $s_{i}^{v}(\beta)$, and that $\overline{x}_{i}^{v}$ instantaneously converges to fixed point $s_{i}$.   The reasoning behind this assumption is that when the patching  update parameter $\alpha$ is small, then the dynamics of patching and filtering update will be on a much slower timescale than the timescale of the malware propagation. This assumption is made in the adaptive control literature \cite{aastrom2013adaptive} for the tractability of analysis. We validated this accuracy of this assumption in Figure \ref{fig:increasing} (b) in Section \ref{sec:simulation}. 

 Under this assumption, we have $\dot{\beta}_{i}(t) = \alpha \sum_{v \in V}{s_{i}^{v}(\beta)}$.  In order to bound the convergence rate, we derive a lower bound on $s_{i}^{v}$ as follows.  We have that
$(1-s_{i})\sum_{v}{\sum_{n_{j} \in N_{i}}{\lambda^{\emptyset,v}s_{j}^{v}}} = \beta_{i}s_{i},$ and hence the union bound $\sum_{v}{s_{j}^{v}} \geq s_{j}$ implies that $(1-s_{i})\sum_{n_{j} \in N_{i}}{\lambda_{min}s_{j}} \leq \beta_{i}s_{i}.$ Summing over $i$ and rearranging terms yields
\begin{IEEEeqnarray*}{rCl}
\sum_{i \in N}{\lambda_{min}|N_{i}|s_{i}} &\leq& 2\sum_{(i,j) \in E}{\lambda_{min}s_{i}s_{j}} + \sum_{i \in N}{\beta_{i}s_{i}} \\
&\leq& \sum_{i \in N}{|N_{i}|\lambda_{min}(s_{i})^{2}} + \sum_{i \in N}{\beta_{i}s_{i}}.
\end{IEEEeqnarray*}
We then arrive at the lower bound $$\sum_{i \in N}{s_{i}(\lambda_{min}|N_{i}|-\beta_{i}-|N_{i}|\lambda_{min}s_{i})} \leq 0.$$ Based on this inequality, we take the approximation $s_{i} \geq \frac{|N_{i}|\lambda_{min} - \beta_{i}}{|N_{i}|\lambda_{min}},$ leading to the dynamics  $$\dot{\beta}_{i}(t) = \alpha\sum_{v \in V}{s_{i}^{v}} \geq \alpha\sum_{v \in V}{\frac{1}{|N_{i}|\lambda_{min}}(|N_{i}|\lambda_{min}-\beta_{i})}.$$ The resulting lower bound on $\beta_{i}(t)$ is then given by $$\beta_{i}(t) \geq \frac{ |V| |N_{i}|}{\overline{\lambda}} + \left(\beta_{i}(0) - \frac{ |V| |N_{i}|}{\overline{\lambda}}\right)\exp{\left(-\alpha\frac{|V|}{|N_{i}|\lambda_{\min}}t\right)}.$$

Next, we consider the final value of $\beta_{i}$ that is reached after the infection rates converge to zero.  The approach is to upper bound $\overline{x}_{i}(t)$, leading to an upper bound on $\dot{\beta}_{i}(t)$ and hence on $\beta_{i}(t)$.  Since $\beta_{i}$ is nondecreasing over time, we have
\begin{IEEEeqnarray*}{rCl}
\dot{\overline{x}}_{i}(t) &\leq& \sum_{j \in N_{i}}{\lambda_{max}(1-\overline{x}_{i})\overline{x}_{j}} - \beta_{i}\overline{x}_{i} \\
&\leq& \lambda_{max}\sum_{j \in N_{i}}{\overline{x}_{j}} - \beta_{i}(0) \overline{x}_{i},
\end{IEEEeqnarray*}
which can be expressed in matrix form as $\dot{\overline{\mathbf{x}}}(t) = (\lambda_{max}A - B_{0})\mathbf{x}(t)$ where $A$ denotes the adjacency matrix of the network and $B_{0}$ is a diagonal matrix with $\beta_{i}(0)$ on the $i$-th diagonal entry.  Hence $$\dot{\overline{\mathbf{x}}}(t) \leq e^{(\lambda_{\max} A - B_{0})t} \overline{\mathbf{x}}(t) \leq e^{(\lambda_{\max} A - B_{0})t}\mathbf{1}.$$  Applying this bound gives
\begin{IEEEeqnarray*}{rCl}
\sum_{i \in N}{\dot{\beta}_{i}(t)} &\leq& \sum_{i \in N}{\sum_{v \in V}{\alpha e^{(\lambda_{\max} A - B_{0})t}\mathbf{1}}} \\
&=& \alpha \mathbf{1}^{T}e^{(\lambda_{\max} A - B_{0})t}\mathbf{1} \leq {\alpha}|N|e^{\mu_{1}(\lambda_{\max} A - B_{0})t},
\end{IEEEeqnarray*}
where $\mu_{1}(\lambda A - B_{0})$ denotes the maximum eigenvalue of the matrix $(\lambda A - B_{0})$.  Integrating yields
\begin{IEEEeqnarray*}{rCl}
&&\sum_{i \in N}{\beta_{i}(t)}-\sum_{i\in N}\beta_{i}(0)\\
&\leq& \frac{|N|\alpha}{|\mu_{1}(\lambda_{\max} A - B_{0})|}(1 - \exp{(-|\mu_{1}(\lambda_{\max} A - B_{0})|t)})
\end{IEEEeqnarray*}
giving a final value $\sum_{i\in N}\beta_{i}^{\ast} \leq \sum_{i\in N}\beta_{i}(0) + \frac{|N|\alpha}{|\mu_{1}(\lambda_{\max} A - B_{0})|}$.

This bound depends on the value of $\beta_{i}(0)$, and is valid whenever $\beta_{i}(0) > \lambda_{max}|N_{i}|$.  We therefore have 
\begin{IEEEeqnarray*}{rCl}
&&\sum_{i \in N}{\beta_{i}^{\ast}} \\
&\leq & \min_{\epsilon_{1},\ldots,\epsilon_{N}}{\left\{\frac{|N|\alpha}{|\mu_{1}(\lambda_{\max} A - B_{0})|} + \sum_{i \in N}{(\lambda_{\max} |N_{i}| + \epsilon_{i})}\right\}},
\end{IEEEeqnarray*}
where $\epsilon_{i} = \beta_{i}(0) - \lambda_{\max} |N_{i}|$. 
 We apply the Greshgorin Circle Theorem \cite{horn2012matrix} to obtain a lower bound $$|\mu_{1}(\lambda_{\max} A - B_{0})| \geq \min{\{\epsilon_{i} : i = 1,\ldots,n\}}.$$
 We have that
\begin{IEEEeqnarray*}{rCl}
\sum_{i \in N}{\beta_{i}^{\ast}} &\leq& \min_{\epsilon}{\left\{\frac{|N|\alpha}{\epsilon} + \lambda_{\max} |E| + |N|\epsilon\right\}} \\
&=& \lambda_{\max} |E| + 2|N|\sqrt{\alpha}. \end{IEEEeqnarray*}
This gives an average $\beta_{i}^{\ast}$ value of approximately $\lambda_{\max} d_{avg} + 2\sqrt{\alpha}$, where $d_{avg}$ is the average degree of the network.

\subsection{Non-Monotone Patching Strategy}
\label{subsec:non-monotone}
The adaptive patching strategy (\ref{eq:adaptive_patching}) results in a patching rate that is monotone nondecreasing in time, and hence may overshoot the malware propagation rate. We now present a patching strategy that can drive the probability of infection to an arbitrarily small final value without exceeding the propagation rate $\lambda$, in the single-virus case. The patching dynamics are defined by 
\begin{equation}
\label{eq:non_monotone_patching}
\dot{\beta}_{i}(t) = \{\alpha x_{i}(t) - \gamma(1-x_{i}(t))\}_{+}.
\end{equation}
This patching strategy can be implemented by incrementing $\beta_{i}(t)$ by $\frac{\alpha}{\beta_{i}(t)}$ when an infection is detected at host ${i}$, and decrementing $\beta_{i}(t)$ by $\frac{\gamma}{\beta_{i}(t)}$ when inspection of host ${i}$ reveals that no virus is present.

Taken together with the single-virus propagation dynamics $$\dot{x}_{i}(t) = \lambda (1-x_{i}(t))\sum_{j \in N_{i}}{x_{j}(t)} - \beta_{i}(t) x_{i}(t),$$ we have that the steady-state values for the infection probability and patching rate are given by $x_{i}^{\ast} = \frac{\gamma}{\alpha + \gamma}$ and $\beta_{i}^{\ast} = \frac{\alpha}{\alpha+\gamma}|N_{i}|\lambda$, respectively. Hence, the probability of infection can be set arbitrarily low, and the patching rate can be set arbitrarily close to the propagation rate, by decreasing $\frac{\gamma}{\alpha}$. 

The local stability of these patching dynamics is governed by the following theorem.
\begin{theorem}
\label{theorem:asymptotic_stability}
The fixed point $(\mathbf{x},\boldsymbol{\beta})$ with $x_{i}^{\ast} = \frac{\gamma}{\alpha + \gamma}$ and $\beta_{i}^{\ast} = \frac{\alpha}{\alpha+\gamma}|N_{i}|\lambda$ for all $i \in N$ is asymptotically stable.
\end{theorem}

\begin{IEEEproof} Linearizing the system around this fixed point, we obtain the Jacobian matrix 
\begin{displaymath}
A = \left(
\begin{array}{cc}
\overline{A} & -\frac{\gamma}{\alpha+\gamma}I \\
(\alpha+\gamma)I & 0
\end{array}
\right),
\end{displaymath}
where $I$ denotes the $|N| \times |N|$ identity matrix and $\overline{A}$ is defined by 
\begin{displaymath}
\overline{A}_{ij} = \left\{
\begin{array}{ll}
-|N_{i}|\lambda, & i=j \\
\frac{\lambda\alpha}{\alpha+\gamma}, & n_{j} \in N_{i} \\
0, & \mbox{else}.
\end{array}
\right.
\end{displaymath}
We  now show that the matrix $A$ is Hurwitz. By Lyapunov's Theorem, a necessary and sufficient condition is to construct a symmetric positive definite matrix $P$ such that $A^{T}P + PA = -\epsilon I$ for some $\epsilon > 0$. 

First, note that $\overline{A}$ is symmetric and is negative definite by the Gershgorin Circle Theorem.
We select a matrix $P$ as 
\begin{displaymath}
P = \left(
\begin{array}{cc}
\tau \overline{A}^{-1} & \frac{(\alpha+\gamma)\epsilon}{2\gamma}I \\
\frac{(\alpha+\gamma)\epsilon}{2\gamma}I & \frac{\gamma}{(\alpha+\gamma)^{2}}\tau\overline{A}^{-1} - \frac{\epsilon}{2\gamma}\overline{A}
\end{array}
\right),
\end{displaymath}
with $\tau = -\frac{1}{2}\left(\epsilon + \frac{(\alpha+\gamma)^{2}\epsilon}{\gamma}\right)$. It can be shown that $A^{T}P + PA = -\epsilon I$. Furthermore, $P$ is symmetric since the matrices $\overline{A}$ and $\overline{A}^{-1}$ are symmetric. It remains to show that $P$ is positive definite. We apply the Schur complement theorem, which states that $P$ is symmetric if and only if
\begin{IEEEeqnarray}{rCl}
\label{eq:non_monotone_proof1}
&&\frac{1}{\tau}\overline{A}^{-1} > 0, \mbox{ and} \\
\label{eq:non_monotone_proof2}
&&\frac{\gamma}{(\alpha+\gamma)^{2}}\tau\overline{A}^{-1} - \left(\frac{\epsilon}{2\gamma} + \left(\frac{(\alpha+\gamma)\epsilon}{2\gamma}\right)^{2}\frac{1}{\tau}\right)\overline{A}  > 0,
\end{IEEEeqnarray}
where ``$>$'' denotes inequality in the positive definite cone. Eq. (\ref{eq:non_monotone_proof1}) holds since $\overline{A}$ is negative definite and $\tau < 0$. After simplifying, eq. (\ref{eq:non_monotone_proof2}) holds  since $\gamma > 0$ and $\overline{A}$ is negative definite. Then there exists a positive definite matrix $P$ such that the Jacobian matrix $A$ satisfies $A^{T}P + PA = -\epsilon I$, implying that the linearized system matrix $A$ is Hurwitz and the fixed point is asymptotically stable.
\end{IEEEproof}

\section{Adaptive Packet Filtering-Based Mitigation}
\label{sec:filtering}
This section presents an adaptive rule for packet filtering-based mitigation. Under the rule, the probability of filtering each packet $q$ is increased with each malware packet that is detected. We first formally define the adaptive filtering-based mitigation strategy, and then analyze the convergence rate and overhead. A joint analysis of patching and filtering-based mitigation is also presented.


\subsection{Adaptive Filtering Strategy}
\label{subsec:filtering_adaptive}
The first step in developing the adaptive filtering strategy is to analyze the passivity of the propagation dynamics when the output is equal to the information available to the packet filtering defense, namely, the rate of packets exchanged between hosts ${i}$ and ${j}$. These passivity properties are analyzed in the following proposition. As a preliminary, define $\overline{\lambda}^{v} = q\mu^{v}$.

\begin{proposition}
\label{prop:H1_passive}
The multi-virus propagation dynamics  are passive from input $((\lambda_{max}^{v}-\overline{\lambda}^{v}) : v \in V)$ to output $(y^{v}(t): (i,j) \in E , v \in V)$, where $$y^{v}(t) = \sum_{(i,j) \in E}{\mu^{v}(\overline{x}_{i}^{v}(1-\overline{x}_{j}^{v}) + \overline{x}_{j}^{v}(1-\overline{x}_{i}^{v}))}$$ and $\mu^{v}$ is the rate at which malware $v$ sends packets to neighboring nodes.
\end{proposition}

\begin{IEEEproof}
Define a storage function by $$W(\mathbf{x}) = \frac{1}{2}\sum_{i=1}^{n}{\sum_{v \in V}{(\overline{x}_{i}^{v})^{2}}}.$$ We then have
\begin{IEEEeqnarray*}{rCl}
\dot{W}(\mathbf{x}) 
 &\leq& \sum_{v \in V}{\sum_{i=1}^{n}{\left[\overline{x}_{i}^{v}\left(\sum_{j \in N(i)}{\lambda^{v}_{max}(1-\underline{x}_{i}^{v})\overline{x}_{j}^{v}}\right.\right.}} \\
 && -\left.\left. \sum_{v \in S}{\sum_{j \in N_{i}}{u_{ij}^{(1)}\mu^{v}x_{i}^{S}(1-\overline{x}_{j}^{v})}}\right)\right],
 \end{IEEEeqnarray*}
where  $\lambda_{max}^{v} = \max{\{\lambda^{S,v} : v \notin S\}}$.  Furthermore, we have that $$\sum_{w \in S}{\mu^{w}x_{i}^{S}(1-\overline{x}_{j}^{w})} \geq \mu^{v}x_{i}^{S}(1-\overline{x}_{j}^{v})$$ for any $v \in S$, and hence
 \begin{multline*}
 \dot{W}(\mathbf{x}) \leq \sum_{v \in V}{\sum_{(i,j) \in E}{\lambda_{max}^{v}(\overline{x}_{i}^{v}\overline{x}_{j}^{v}(1-\overline{x}_{i}^{v}) + \overline{x}_{i}^{v}\overline{x}_{j}^{v}(1-\overline{x}_{j}^{v}))}} \\
  - \sum_{v \in V}{\sum_{(i,j) \in E}{q\mu^{v}((\overline{x}_{i}^{v})^{2}(1-\overline{x}_{j}^{v}) + (\overline{x}_{j}^{v})^{2}(1-\overline{x}_{i}^{v}))}}.
  \end{multline*}
  Now, since $2\overline{x}_{i}^{v}\overline{x}_{j}^{v} \leq (\overline{x}_{i}^{v})^{2} + (\overline{x}_{j}^{v})^{2}$, we have that
  \begin{multline*}
  \dot{W}(\mathbf{x}) \leq \sum_{v \in V}{\sum_{(i,j) \in E}{(\lambda_{max}^{v}-\overline{\lambda}^{v})\left((\overline{x}_{i}^{v})^{2}(1-\overline{x}_{j}^{v})  \right.}}  \\ + \left.(\overline{x}_{j}^{v})^{2}(1-\overline{x}_{i}^{v})\right).
  \end{multline*}
  This completes the proof of passivity.
\end{IEEEproof}

Proposition \ref{prop:H1_passive} implies that the propagation dynamics are passive from input $(\lambda-\overline{\lambda})$ to output $y^{v}$.  We consider the filtering probability update rule 
\begin{equation}
\label{eq:q_dynamics_new}
\dot{q}(t) = \gamma\left\{\sum_{(i,j) \in E}{\sum_{v \in V}{\mu^{v}(\overline{x}_{i}^{v}(1-\overline{x}_{j}^{v}) + \overline{x}_{j}^{v}(1-\overline{x}_{i}^{v}))}}\right\}_{q < 1},
\end{equation}
where $\{f(\mathbf{x})\}_{q<1} = f(\mathbf{x})$ if $q < 1$ and $0$ otherwise.
This update rule can be implemented by incrementing $q(t)$ by $\frac{\gamma}{q(t)}$ whenever a malware packet is detected, since $q(t)$ and $\gamma$ are known parameters at each time $t$. To show that this update rule results in (\ref{eq:q_dynamics_new}), we observe that the rate of the filtering update process is given as 
$$q(t)\left(\sum_{(i,j) \in E}{\sum_{v \in V}{\mu^{v}(\overline{x}_{i}^{v}(1-\overline{x}_{j}^{v}) + \overline{x}_{j}^{v}(1-\overline{x}_{i}^{v}))}}\right).$$ Therefore, by the same logic as the derivation of (\ref{eq:adaptive_patching}), incrementing the filtering probability by $\frac{\gamma}{q}$ when $q < 1$ results in the dynamics (\ref{eq:q_dynamics_new}), which does not require the knowledge of the propagation rate $\lambda^{v}$.


\begin{theorem}
\label{theorem:still_converges}
The update rule (\ref{eq:q_dynamics_new}) guarantees convergence of $\overline{x}_{i}^{v}$ to $0$ for all $i \in N$ and $v \in V$.
\end{theorem}

\begin{IEEEproof}
Define a storage function $W(\mathbf{x},q)$ by $$W(\mathbf{x},q) = \left\{
\begin{array}{ll}
\frac{1}{2}\sum_{i \in N}{\sum_{v \in V}{(\overline{x}_{i}^{v})^{2}}} & \\
 + \frac{1}{2}(q-\bar{p})^{2}, & q < \bar{p} \\
\frac{1}{2}\sum_{i \in N}{\sum_{v \in V}{(\overline{x}_{i}^{v})^{2}}}, & q \geq \bar{p}.
\end{array}
\right.$$
Then $\dot{W}(\mathbf{x},q)$ is bounded by
\begin{IEEEeqnarray*}{rCl}
\IEEEeqnarraymulticol{3}{l}{
\dot{W}(\mathbf{x},q)} \\
 &\leq& \sum_{v \in V}{\sum_{(i,j) \in E}{\mu^{v}(p^{v}-q)((1-\overline{x}_{i}^{v})(\overline{x}_{j}^{v})^{2} + (1-\overline{x}_{j}^{v})(\overline{x}_{i}^{v})^{2})}} \\
&& + \sum_{v \in V}{\sum_{(i,j) \in E}{\mu^{v}(q-p^{v})((1-\overline{x}_{i}^{v})\overline{x}_{j}^{v} + (1-\overline{x}_{j}^{v})\overline{x}_{i}^{v})}} \\
&&  - \sum_{i \in N}{\sum_{v \in V}{\beta_{i}(\overline{x}_{i}^{v})^{2}}} \leq - \sum_{i \in N}{\sum_{v \in V}{\beta_{i}(\overline{x}_{i}^{v})^{2}}} < 0
\end{IEEEeqnarray*}
when $q < \bar{p}$ and $\dot{W}(\mathbf{x},q) < 0$ when $q \geq \bar{p}$ as well.  Hence, the function $W$ is strictly decreasing and converges to the set $\{\dot{W} = 0\}$, which occurs exactly when $\overline{x}_{i}^{v} = 0$ for all $i \in N$ and $v \in V$.
\end{IEEEproof}

\subsection{Convergence Rate Analysis}
\label{subsec:convergence}
The convergence rate of the filtering probability to a sufficiently large value will determine how quickly the network defense is able to mitigate the malware propagation. In order to analyze the convergence rate, we divide the time required for all viruses to be removed into two intervals.  The first time interval is the time for $q(t)$ to increase until it approaches $p^{v}_{max}$; this can be interpreted as the time required to ``learn'' the correct filtering strategy.  The second time interval is the time required for all viruses to be removed after $q(t)$ has reached this threshold value.

For simplicity, we define $\underline{\beta} = \min_{i\in N} \beta_{i}$, $\bar{\beta} = \max_{i\in N} \beta_{i}$ and $\bar{p} = \max_{S,v} p^{S,v}$, $\underline{p} = \min_{S,v} p^{S,v}$. Similarly $\bar{p}^{v} = \max_{S} p^{S,v}$ and $\underline{p}^{v} = \min_{S} p^{S,v}$. 
We analyze the time required for $q(t)$ to approach $(p^{v}_{max}-\beta_{i})$.  Let $\{r_{i}^{v}(q) : i \in N, v \in V\}$ denote a fixed point of $\overline{x}_{i}^{v}$ when $q(t)$ is constant and equal to $q$; when $q$ is small, there exists such a fixed point with $r_{i}^{v} > 0$ for all $i \in N$ and $v \in V$.  In order to analyze the convergence rate of $q(t)$, we adopt an approximation where the dynamics of $\overline{x}_{i}^{v}$ converge instantaneously to $r_{i}^{v}$ for all $i$ and $v$.     

Under this approximation, the  dynamics of $q(t)$ are 
\begin{multline}
\label{eq:q_dynamics_approx}
\dot{q}(t) \approx \gamma \{\sum_{v \in V}{\sum_{(i,j) \in E}{\mu^{v}\left((1-r_{i}^{v}(q))r_{j}^{v}(q)  \right. }}\\
\left. \left. + (1-r_{j}^{v}(q))r_{i}^{v}(q)\right)\right\}_{q < 1}.
\end{multline}
A lower bound on the convergence time is described as follows.

\begin{proposition}
\label{prop:filtering_convergence_time}
The filtering probability $q(t)$ satisfies 
\begin{equation}
\label{eq:q_lower_bound}
\dot{q}(t) \leq \frac{\gamma |V|\bar{\beta}}{\underline{p}-q}\left( \min_{i\in N}|N_{i}| + \frac{(|N|-\min_{i\in N}|N_{i}|)\lambda_{max}-\underline{\beta}}{\mu_{min}(\underline{p}-q)}\right)
\end{equation}
 when $q(t) \leq \underline{p}$.
\end{proposition}

\begin{IEEEproof}
Denote $m = |V|$, $n = |N|$ and $d_{min} = \min_{i\in N} |N_{i}|$. 
We have that $$\dot{q}(t) = \sum_{v \in V}{\gamma\sum_{(i,j) \in E}{\mu^{v}(r_{i}^{v}(1-r_{j}^{v}) + r_{j}^{v}(1-r_{i}^{v}))}},$$ which can be bounded as
\begin{IEEEeqnarray*}{rCl}
\dot{q}(t) &=& \sum_{v \in V}{\left[\frac{\gamma}{\underline{p}^{v}-q}\sum_{(i,j) \in E}{\mu_{v}(\underline{p}^{v}-q)(r_{i}^{v}(1-r_{j}^{v}) + r_{j}^{v}(1-r_{i}^{v}))}\right]} \\
&\leq& \sum_{v \in V}{\left[\frac{\gamma}{\underline{p}^{v}-q}\sum_{(i,j) \in E}{\left(\sum_{S: v \notin S}{p^{S,v}\mu^{v}(r_{i}^{S}(1-r_{j}^{v}) + r_{j}^{S}(1-r_{i}^{v}))}\right)}\right]} \\
\label{eq:filtering_fixed_pt}
&=& \sum_{v \in V}{\frac{\gamma}{\underline{p}^{v}-q}\sum_{i \in N}{\beta_{i}r_{i}^{v}}},
\end{IEEEeqnarray*}
where (\ref{eq:filtering_fixed_pt}) follows from the fact that $r_{i}^{v}$ is a fixed point of the dynamics of $\overline{x}_{i}^{v}$.

Next, an upper bound on $\sum_{i \in N}{\beta_{i}r_{i}^{v}}$ is derived. At the fixed point, 
\begin{IEEEeqnarray*}{rCl}
\beta_{i}r_{i}^{v} &=& \sum_{S: v \notin S}{\sum_{j \in N_{i}}{\lambda^{S,v}r_{i}^{S}r_{j}^{v}}} - \sum_{S \ni v}{\sum_{w \in S}{\sum_{j \in N_{i}}{q\mu^{w}r_{i}^{S}(1-r_{j}^{w})}}} \\
&\leq& \lambda_{max}^{v}(1-r_{i}^{v})r_{j}^{v} - \sum_{S \ni v}{\sum_{w \in S}{\sum_{j \in N_{i}}{q\mu^{w}r_{i}^{S}(1-r_{j}^{w})}}} \\
&\leq& (1-r_{i}^{v})\lambda_{max}^{v}\sum_{j \in N_{i}}{r_{j}^{v}} - q\mu^{v}r_{i}^{v}\sum_{j \in N_{i}}{(1-r_{j}^{v})} 
\end{IEEEeqnarray*}
Rearranging terms yields 
\begin{IEEEeqnarray*}{rCl}
r_{i}^{v} &\leq& \frac{\lambda_{max}^{v}\sum_{j \in N_{i}}{r_{j}^{v}}}{\beta_{i} + \lambda_{max}^{v}\sum_{j \in N_{i}}{r_{j}^{v}} + q\mu^{v}\sum_{j \in N_{i}}{(1-r_{j}^{v})}} \\
&=& \frac{\lambda_{max}^{v}\sum_{j \in N_{i}}{r_{j}^{v}}}{\beta_{i} + (\lambda_{max}^{v}-q\mu^{v})\sum_{j \in N_{i}}{r_{j}^{v}} + q\mu^{v}|N_{i}|} \\
&\leq& \frac{\lambda_{max}^{v}\sum_{j \in N_{i}}{r_{j}^{v}}}{\beta_{i} + (\lambda_{max}^{v}-q\mu^{v})\sum_{j \in N_{i}}{r_{j}^{v}} + q\mu^{v}|N_{i}|}
\end{IEEEeqnarray*}
 Summing over $i$ then gives
 $$\sum_{i \in N}{r_{i}^{v}} \leq \frac{n\lambda_{max}^{v} - (\underline{\beta} + q\mu^{v}d_{min})}{\lambda_{max}^{v}-q\mu^{v}}.$$ Combining with (\ref{eq:filtering_fixed_pt}), we have 
 \begin{IEEEeqnarray*}{rCl}
 \dot{q}(t) &\leq& \sum_{v \in V}{\left[\frac{\gamma \bar{\beta}}{\underline{p}^{v}-q} \cdot \frac{n \lambda_{max}^{v}-(\underline{\beta} + q\mu^{v}d_{min})}{\mu^{v}\bar{p}^{v}-q\mu^{v}}\right]} \\
 &\leq& \sum_{v \in V}{\left[\frac{\gamma \bar{\beta}}{\underline{p}^{v}-q} \cdot \frac{n \lambda_{max}^{v}-(\underline{\beta} + q\mu^{v}d_{min})}{\mu^{v}\underline{p}^{v}-q\mu^{v}}\right]} \\
 &\leq& \frac{\gamma\bar{\beta}m}{\underline{p}-q}\left(d_{min} + \frac{(n-d_{min})\lambda_{max} - \underline{\beta}}{\mu_{min}(\underline{p}-q)}\right),
 \end{IEEEeqnarray*}
 completing the proof.
\end{IEEEproof}

The upper bound on $q(t)$ can be used to analyze the time required for the filtering probability to converge to $\underline{p}$. 

%
%
%
%
%

%

\subsection{Final Value of Filtering Probability}
\label{subsec:filtering_analysis}
The filtering probability $q(t)$ is a monotone increasing function that is bounded above by $1$, and hence converges to a value $q^{\ast} = \lim_{t \rightarrow \infty}{q(t)}$.  If this final value is approximately equal to $\bar{p}$, then the network will filter just enough packets to ensure that all viruses are removed.  On the other hand, if $q^{\ast}$ is approximately equal to $1$, then almost all packets (including non-malware packets) will be inspected, increasing the delays experienced by legitimate network traffic.  In what follows, we analyze the value of $q^{\ast}$ as a function of the parameters $\gamma$ and $\beta$.

\begin{proposition}
\label{prop:q_overshoot}
The final value of $q(t)$ satisfies
\begin{equation}
\label{eq:q_ast_bound}
q^{\ast} \leq \min{\left\{\bar{p} + |V|\gamma\sum_{i \in N}{\frac{|N_{i}|}{\beta_{i}}},1\right\}}.
\end{equation}
\end{proposition}

\begin{IEEEproof}
By inspection of (\ref{eq:q_dynamics_new}) and the fact that $(1-\overline{x}_{i}^{v}) \leq 1$, we have that
\begin{IEEEeqnarray}{rCl}
\dot{q}(t) &\leq& \gamma\left\{\sum_{v \in V}{\sum_{(i,j) \in E}{(\overline{x}_{i}^{v} + \overline{x}_{j}^{v})}}\right\}_{q < 1} \\
&=& \gamma\left\{\sum_{v \in V}{\sum_{i \in N}{|N_{i}|\overline{x}_{i}^{v}}}\right\}_{q < 1} \\
\label{eq:overshoot1}
&\leq& \gamma\left\{\sum_{v \in V}{\sum_{i \in N}{|N_{i}|e^{-\beta_{i}t}}}\right\},
\end{IEEEeqnarray}
where (\ref{eq:overshoot1}) follows from the upper bound $\overline{x}_{i}^{v}(t) \leq e^{-\beta_{i}t}$ when $q > \bar{p}$.  This yields $$q(t) \leq q(0) + \sum_{v \in V}{\sum_{i \in N}{\frac{|N_{i}|\gamma}{\beta_{i}}\left(1 - e^{-\beta_{i}t}\right)}}.$$ The expression can be simplified by noting that $q(0) = \bar{p}$ and the inner summation of the second term has no dependence on $v$.  The fact that $\dot{q}(t) = 0$ when $q = 1$ then implies (\ref{eq:q_ast_bound}).
\end{IEEEproof}

The following Corollary shows that when both adaptive patching and filtering are employed, all malwares are removed independent of the initial values of $\beta_{i}(0), q(0)$ and update parameters $\alpha, \gamma$ as long as these parameters are positive. 
\begin{corollary} Joint adaptive patching and filtering guarantee convergence of  $\bar{x}_{i} = 0$ for all $i\in N$ for any $q(0)>0$, and $\beta_{i}(0)>0$. 
\end{corollary}
\begin{IEEEproof}
Define the dynamics of $\bar{x}_{i}$ with only adaptive patching as $\dot{\bar{x}}_{i}^{(p)}$ and the dynamics of joint adaptive patching and filtering as $\dot{\bar{x}}_{i}^{(p), (f)}$. Since 
\begin{equation*}
\dot{\bar{x}}_{i}^{(p), (f)} = \dot{\bar{x}}_{i}^{(p)} - \sum_{v\in V} q(t)\mu^{v} \bar{x}_{i}^{v}\sum_{j\in N_{i}} (1-\bar{x}_{j}^{v}),
\end{equation*}
and $\sum_{v\in V} q(t)\mu^{v} \bar{x}_{i}^{v}\sum_{j\in N_{i}} (1-\bar{x}_{j}^{v}) \geq 0$ for all $\bar{x}_{i}^{v}$ and $q(t)>0$, we have $\dot{\bar{x}}_{i}^{(p), (f)}\leq \dot{\bar{x}}_{i}^{(p)}$. Since $q(t) > 0$ for all $t$ for $q(0)>0$ since $q(t)$ is a monotonic non-decreasing function in $t$. Therefore, for the same initial point $\bar{x}_{i}(0)$, the trajectory of $\bar{x}_{i}^{(p),(f)}(t)$ with joint patching and filtering will be upper bounded by the trajectory of $\bar{x}_{i}^{(p)}(t)$ with only filtering. However, Theorem \ref{theorem:patching_stable} shows that under $\dot{x}_{i}^{(p)}$, $\bar{x}_{i}(t)$ will converge to 0 for all initial points $\bar{x}_{i}(0) \in [0,1]$. Therefore, the joint adaptive patching and filtering guarantee convergence to $\bar{x}_{i} = 0$.
\end{IEEEproof}

\section{Simulation Study}
\label{sec:simulation}

In this section, we conduct a numerical study via MatlabTM. We conduct three numerical studies. First, we compare the mean-field approximation with the underlying Markov process by comparing the trajectories in the static patching case. Second, we simulate the adaptive patching strategy where the patching rate for host $i$ is incrementally increased when the infection of host $i$ is detected, as well as the adaptive filtering strategy jointly employed with static patching. Finally, we conduct a numerical study for the non-monotonic increasing adaptive patching strategy proposed in Section \ref{subsec:non-monotone}.

\begin{figure}[h]
$\begin{array}{cc}
\includegraphics[width=1.9in]{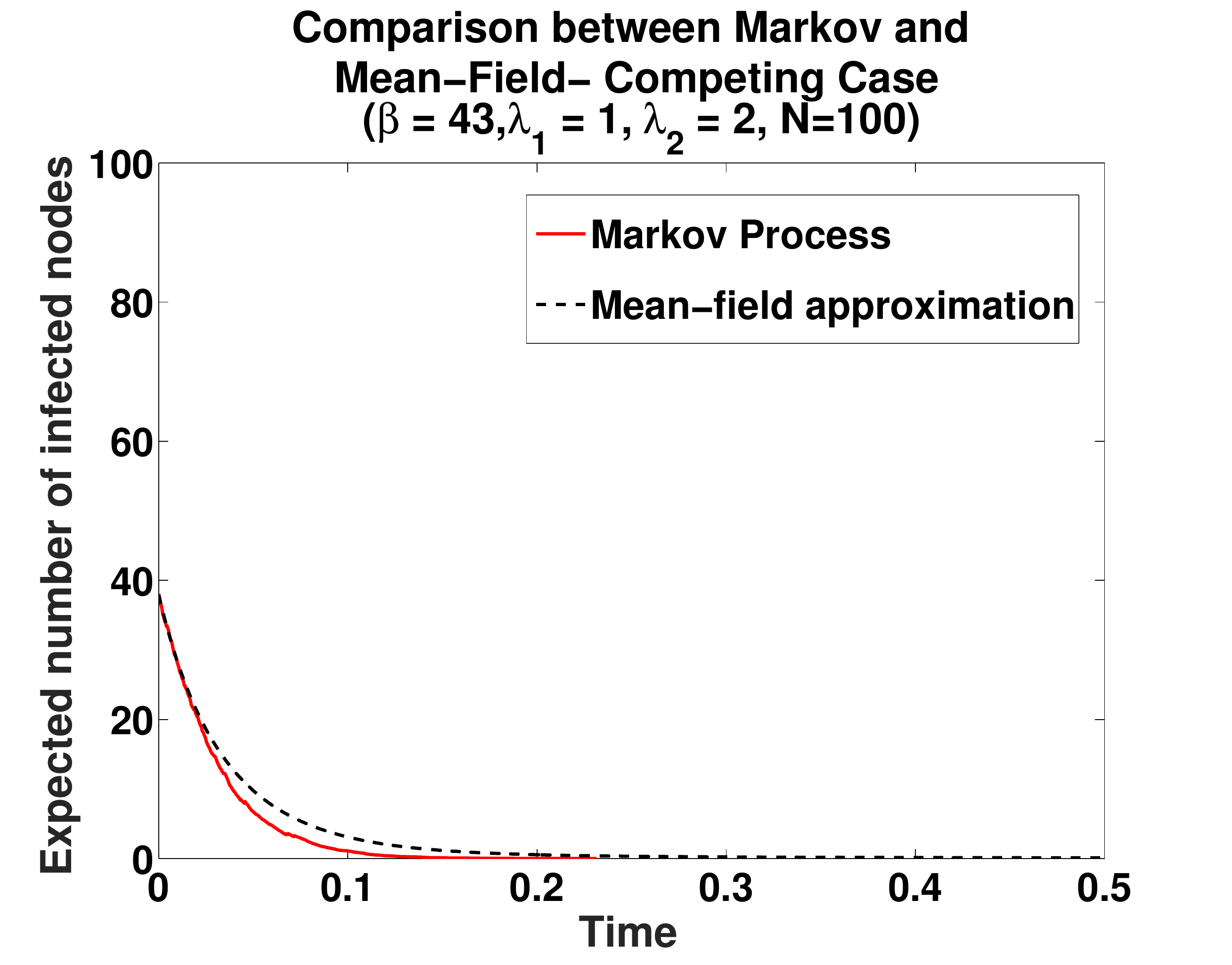} &
\includegraphics[width =1.7in]{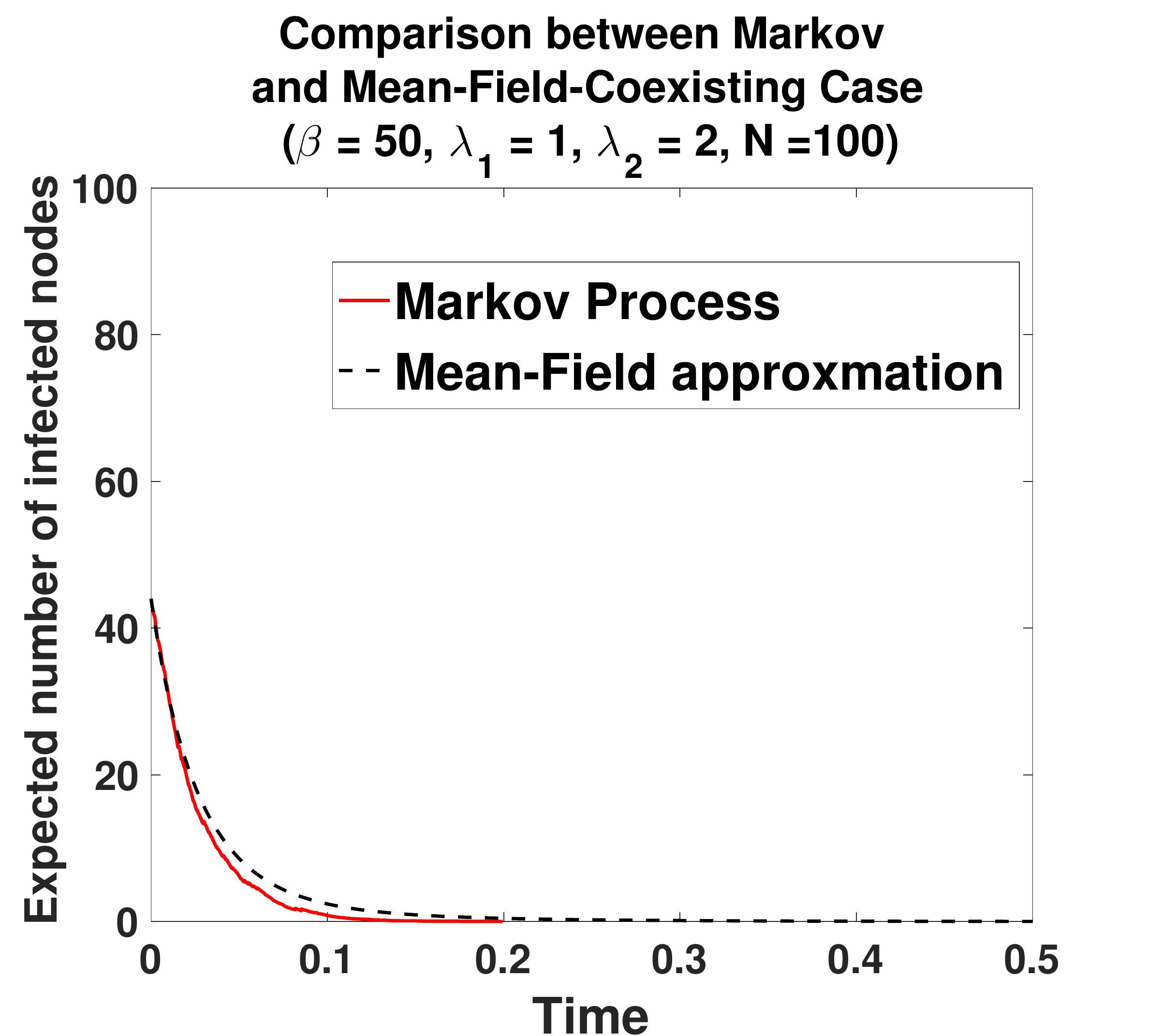} \\
\mbox{(a)} & \mbox{(b)}\\
\end{array}$
\caption{Figure comparing the Markov process and the mean-field approximation with independence assumption. In both competing and coexisting cases, mean-field approximation provide good approximation while providing upper bounds on the trajectory of $\bar{x}_{i}(t)$ which is consistent with Theorem \ref{thm:xbar}.}
\label{fig:markov}
\end{figure}

We assume there are two viruses $v_{1}, v_{2}$ propagating through the network, and the infection rates are given as $\lambda^{S, \{v_{1}\}} = \lambda_{1} = 1$ and $\lambda^{S,  \{v_{2}\}} = \lambda_{2} = 2$ for all sets $S \subset \{v_{1}, v_{2}\}$ in the coexisting case, and the same infection rates are given as $\lambda^{\emptyset, v_{1}} = \lambda^{v_{2}, v_{1}} = 1$, $\lambda^{\emptyset, v_{2}} = \lambda^{v_{1}, v_{2}} = 2$ in the competing case. For the comparison between Markov process and mean-field approximation, we considered a Erdos-Renyi graph with 100 hosts and probability of connection $p=0.2$. We assume that initially, each host is infected with either malware 1 or 2 with probability $0.4$. To simulate the underlying Markov process, we used Monte-Carlo methods with 100 trials. Figure \ref{fig:markov} validates that mean-field approximation provide good approximation of the underlying Markov chain for both the competing and coexisting cases. It also shows that mean-field approximation with independence assumption provides an upper bound on the trajectories of $\bar{x}_{i}(t)$ as proved in Theorem \ref{thm:xbar}. Figure \ref{fig:markov} also illustrates that when $\beta$ values are chosen to satisfy the passivity index conditions shown in Theorem \ref{theorem:SIS_passive_2}, it is sufficient to remove all malwares at desired rates.

%

\begin{figure}[h]
\centering
$\begin{array}{cc}
\includegraphics[width=1.8in]{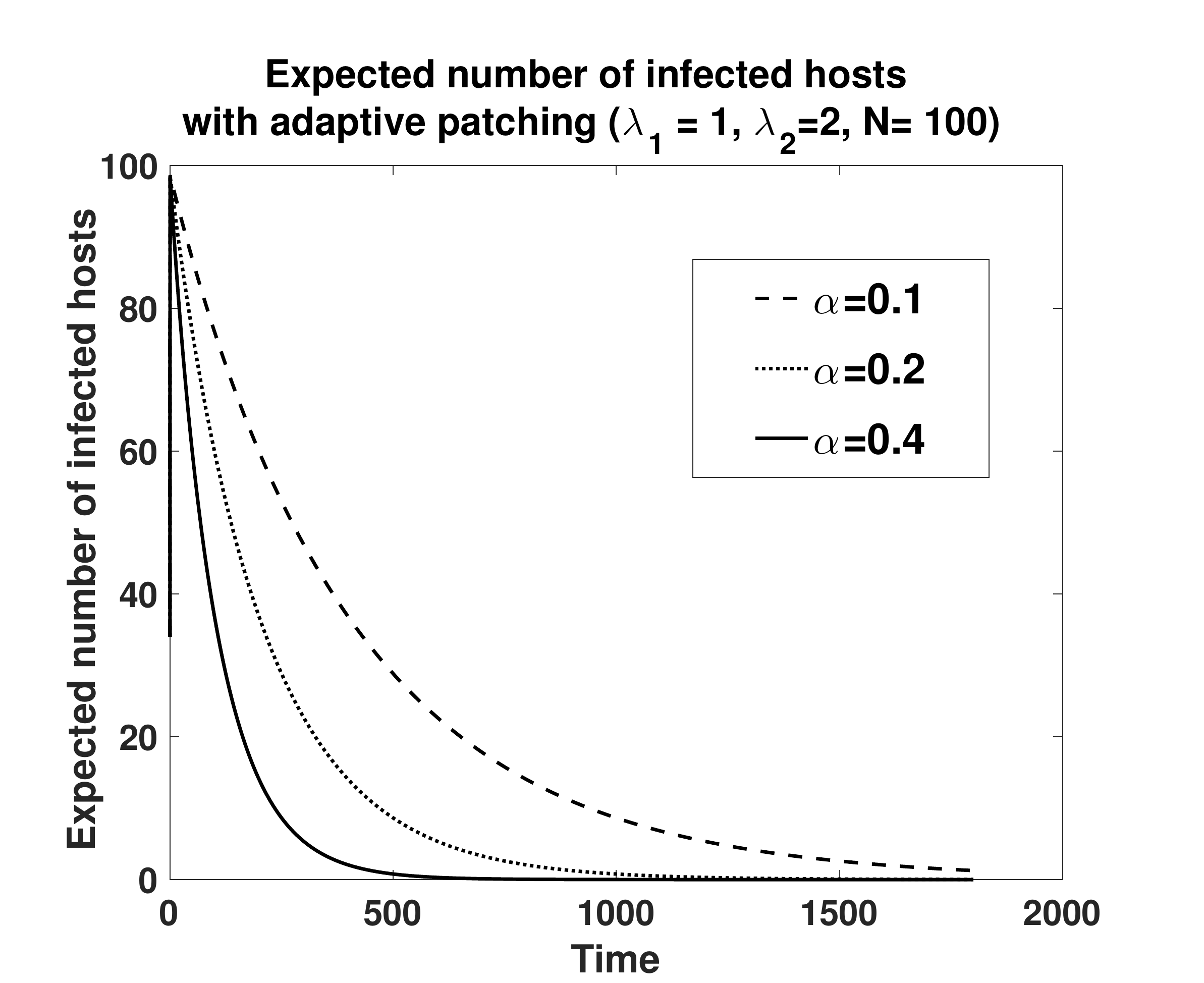} &
\includegraphics[width =1.7in]{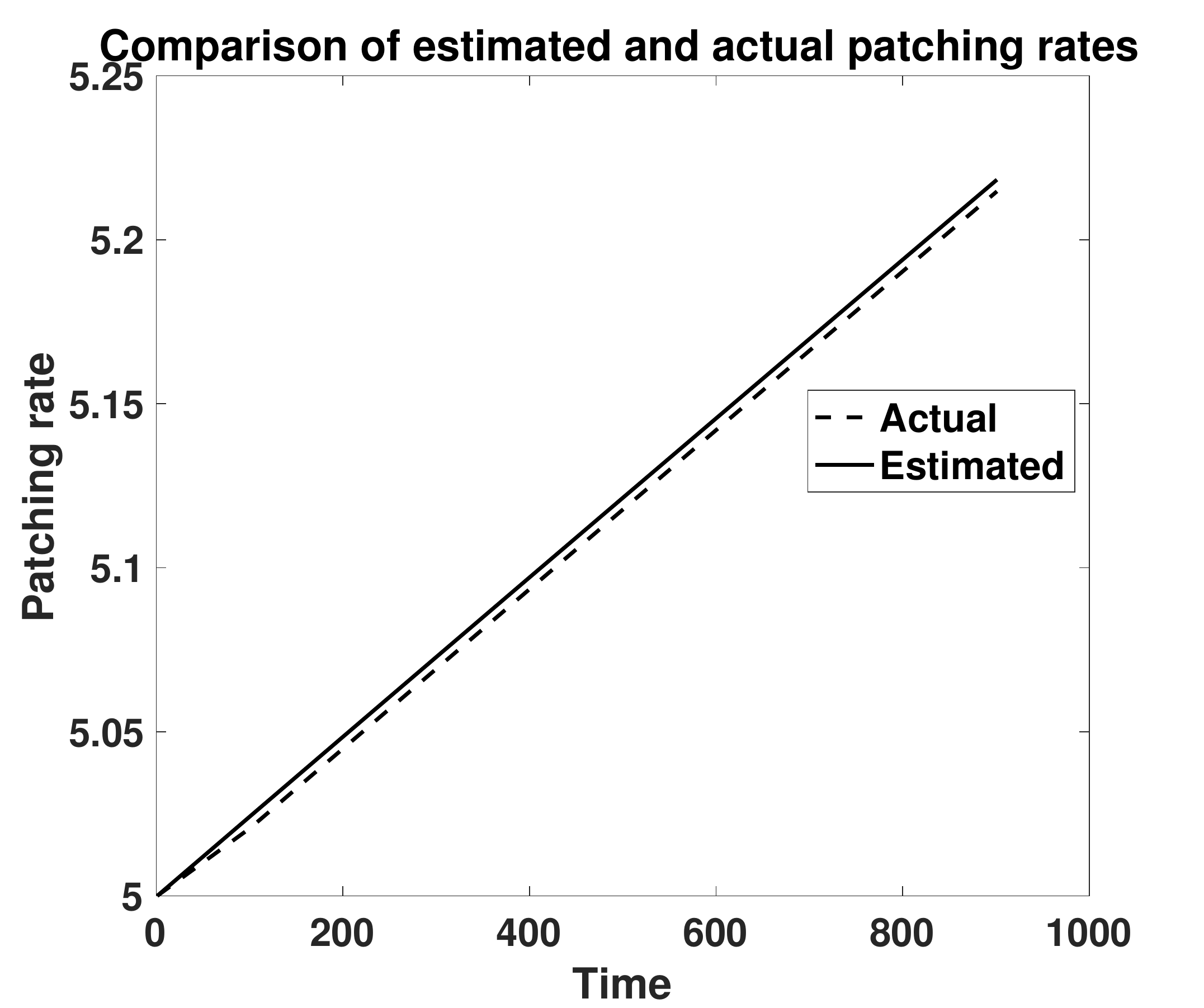} \\
\mbox{(a)} & \mbox{(b)}\\
\end{array}$
\caption{(a): illustration of the effectiveness of adaptive patching strategy. Higher values of $\alpha$ ensures faster convergence rate to the final value at the cost of higher final patching rates at the equilibrium. (b) Comparison between the estimated patching rate with the instantaneous convergence assumption in Section \ref{subsec:adaptive_patching} and the actual trajectory of $\beta$.}
\label{fig:increasing}
\end{figure}

The convergence of patching rates for the non-decreasing adaptive patching strategy for different $\alpha$ values are shown in Figure \ref{fig:increasing} (a) for both competing and coexisting cases. The network configuration is same as the static patching rate case, and the initial $\beta$ values were set to 10 for all hosts.  Figure \ref{fig:increasing} (b) validates the assumption of the instantaneous convergence to the fixed point in Section \ref{subsec:adaptive_patching}. The errors introduced by the instantaneous convergence assumption is negligible from the actual trajectory $\beta_{i}(t)$.

\begin{figure}[h]
$\begin{array}{cc}
\includegraphics[width=1.70in]{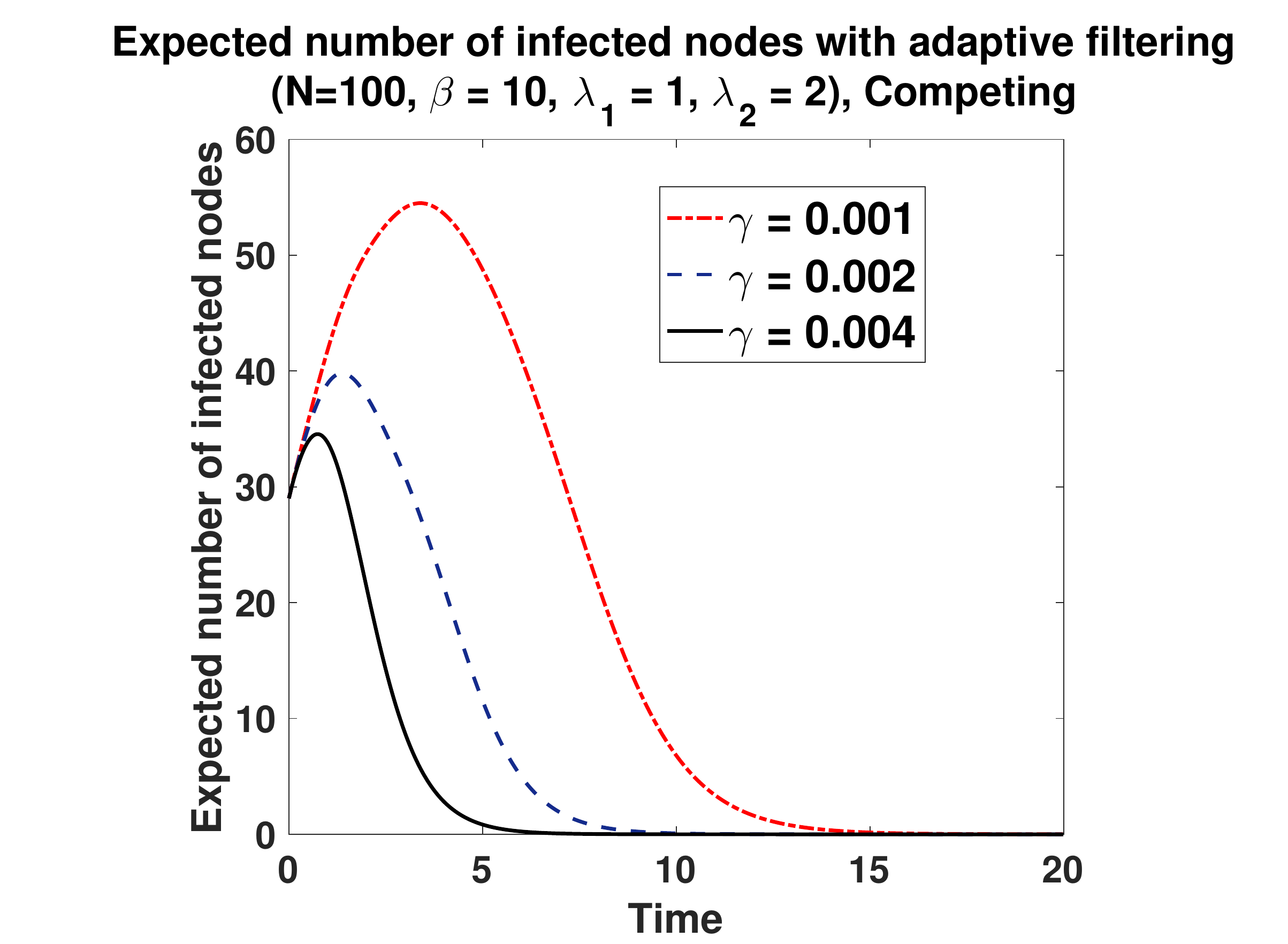}&
\includegraphics[width=1.7in]{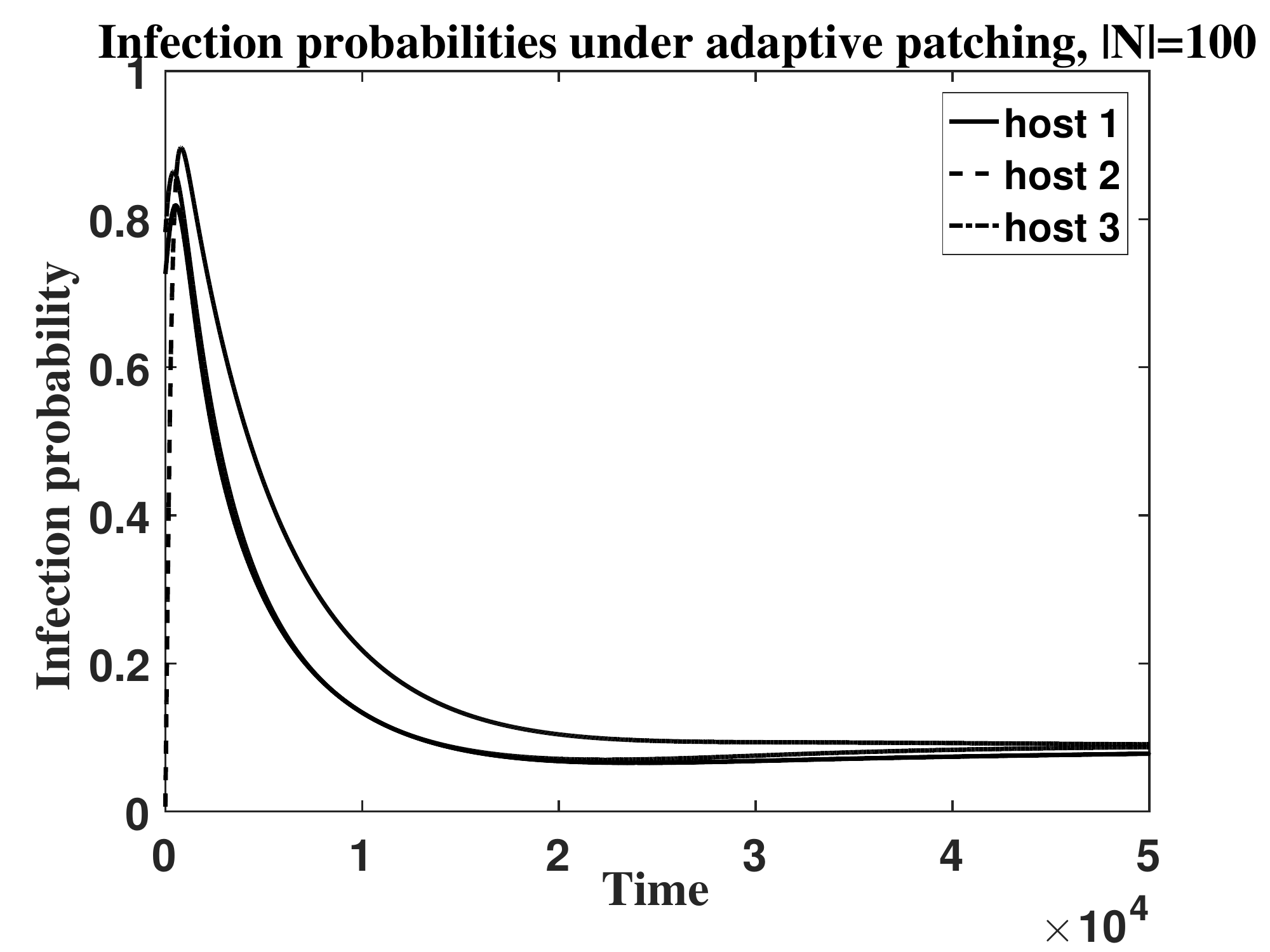} \\
\mbox{(a)} & \mbox{(b)}\\
\end{array}$
\caption{(a) Figure illustrating the effectiveness of adaptive filtering strategy. Adaptive filtering strategy is employed jointly with a static patching strategy with rate $\beta_{i} = 10$ for all hosts. Smaller values of $\gamma$ results lower final values of $q$ at the cost of  higher peak number of infected hosts and longer time till all malwares are removed. (b) Effectiveness of non-monotone patching strategy. Probability of infection asymptotically converges to the equilibrium point computed in Theorem \ref{theorem:asymptotic_stability}.}
\label{fig:adaptivefiltering}
\end{figure}

The effectiveness of the adaptive filtering strategy with static patching rate of $\beta_{i}=10$ for all hosts are illustrated in Figure \ref{fig:adaptivefiltering}. Propagation rates $\lambda_{1}$, $\lambda_{2}$ are same as the static patching rate and the network was chosen to be a Erdos-Renyi random graph with $p=0.2$. Initially, each host is infected with either virus 1 or 2 with probability 0.3. Figure \ref{fig:adaptivefiltering} (a) shows that all malwares are eventually removed from the network. Smaller update parameter $\gamma$ results in low final values of $q(t)$ (Figure \ref{fig:adaptivefiltering} (b)) at the cost of longer time to remove all malwares. 

Figure \ref{fig:adaptivefiltering} (a) verifies that the adaptive patching strategy in Section \ref{subsec:adaptive_patching} removes all malwares from the network. Large update parameter $\alpha$ ensures faster convergence to the desired steady state at the cost of higher final average patching rate at the equilibrium, resulting in unnecessarily high patching rates. 

The non-monotone adaptive patching rule (Figure \ref{fig:adaptivefiltering} (b)) was evaluated as follows. We considered propagation of a single virus in an Erdos-Renyi random graph with $100$ hosts and $p=0.05$. The propagation rate was $\lambda = 1$, while $\alpha = 1$ and $\gamma = 0.1$. For each host, the initial infection probabilities and patching rates were chosen independently and uniformly at random from $[0,1]$ and $[0,0.2]$, respectively. The trajectory of $x_{i}(t)$ for $i=1,2,3$ is shown in Figure \ref{fig:adaptivefiltering}(b). Each of the three trajectories converges to the fixed point $\frac{\gamma}{\alpha+\gamma}$ from the initial state. We observed this behavior in all independent trials that were run, leading us to conjecture that convergence to the desired steady-state occurs from any initial state and hence is not a purely local phenomenon.

\section{Conclusions}
\label{sec:conclusion}
In this paper, we investigated static and adaptive mitigation strategies against propagation of multiple competing and coexisting malwares. We developed a passivity-based framework, and proved that patching and filtering-based defenses can be analyzed and designed jointly by modeling them as coupled dynamical systems. In the case where the malware propagation rates are known \emph{a priori}, we characterized the needed patching rate as a passivity index of the dynamical model. We formulated the problem of selecting the minimum-cost mitigation strategy to remove all viruses at a desired rate by leveraging the derived passivity index.

When the propagation rates are not known \emph{a priori}, we presented adaptive mitigation strategies that vary the rate of patching a host, or the probability of filtering a packet, in response to the observed malware infections. We developed two adaptive patching strategies, namely, a monotone increasing patching rate that guarantees removal of all viruses in steady-state, as well as a non-monotone patching rate that can approximate the propagation rate to any desired accuracy by varying the mitigation parameters. We also presented an adaptive packet filtering strategy for removing all viruses. 

The adaptive update strategies presented in this paper involve each host updating its own patching rate based on its observed infection probability. In future work, We will investigate generalizations to other propagation models, such as Susceptible-Infected-Recovered. Also, while we showed that joint adaptive patching and filtering remove all malwares, finding the optimal tradeoff between two mitigation strategies by tuning the update parameters is an open research problem.

\bibliographystyle{IEEEtran}
\bibliography{Poovendran_TCNS_2016_malware.bib}

\end{document}